\documentclass[10pt,twocolumn,twoside]{IEEEtran}
\IEEEoverridecommandlockouts
\usepackage{cite}
\usepackage{amsmath,amssymb,amsfonts}
\usepackage{bm}
\usepackage{microtype}
\usepackage{graphicx}
\usepackage{textcomp}
\usepackage{xcolor,float}
\usepackage{enumitem} 
\usepackage{comment}
\usepackage{threeparttable}
\usepackage[utf8]{inputenc}
\usepackage[ruled, vlined]{algorithm2e}
\usepackage{booktabs}
\usepackage[hidelinks]{hyperref}

\allowdisplaybreaks
\SetInd{0.5em}{0.5em}
\SetKwBlock{primaldual}{\textnormal{Projected primal-dual for variables:}}{end}
\SetKwBlock{dynamictracking}{\textnormal{Dynamic tracking of the aggregate:}}{end}
\SetKwBlock{OuterLoop}{For $k=0,1,\dots$ do}{}

\makeatletter

\newcommand{\Rmnum}[1]{\expandafter\@slowromancap\romannumeral #1@}
\makeatother

\newtheorem{theorem}{Theorem}

\newtheorem{assumption}{Assumption}
\newtheorem{remark}{Remark}
\newtheorem{definition}{Definition}
\newtheorem{prop}{Proposition}

\def\BibTeX{{\rm B\kern-.05em{\sc i\kern-.025em b}\kern-.08em
    T\kern-.1667em\lower.7ex\hbox{E}\kern-.125emX}}
\begin{document}

\title{Distributed \!Algorithm \!for\! Robust\! Wardrop Equilibrium \\in Uncertain Aggregative Congestion Games\\

\thanks{This work was supported by the Swedish Research Council Distinguished
Professor Grant 2017-01078, the Knut and Alice Wallenberg Foundation
Wallenberg Scholar Grant, the Wallenberg AI, Autonomous Systems and Software Program (WASP) funded by the Knut and Alice Wallenberg Foundation, and the Swedish Strategic Research Foundation
SUCCESS Grant FUS21-0026.}
%\textit{Corresponding author: G. Chen.}}
\thanks{
H. Peng, G. Belgioioso and K. H. Johansson are with School of Electrical Engineering and Computer Science, KTH Royal Institute of Technology, 100 44 Stockholm, Sweden, and also with Digital Futures, 100 44 Stockholm, Sweden (e-mail: \{huanp, giubel, kallej\}@kth.se).}
\thanks{
G. Chen is with School of Automation, Southeast University, Nanjing 210096, China (e-mail: guanpu\_chen@seu.edu.cn).}
}

\author{\IEEEauthorblockN{Huan Peng,}
\and
\IEEEauthorblockN{Guanpu Chen,}
\and
\IEEEauthorblockN{Giuseppe Belgioioso,}
\IEEEauthorblockN{Karl Henrik Johansson, \textit{Fellow, IEEE}}
%\vspace{-5pt}
}

\maketitle
\begin{abstract}
    This paper considers a class of aggregative congestion games with uncertain coupling constraints, and devises a distributed algorithm to seek the robust generalized Wardrop equilibrium (RGWE) under worst-case uncertainty. Utilizing robust optimization theory, we reformulate the original aggregative congestion game with uncertainty into a tractable and deterministic augmented problem. Building upon this reformulation, we design a fully distributed algorithm to seek the RGWE by integrating a projected primal-dual scheme and a dynamic tracking technique. The convergence of the proposed algorithm is rigorously guaranteed via singular perturbation theory and LaSalle's invariance principle. Furthermore, we explicitly characterize the relationship between the obtained RGWE and the robust generalized Nash equilibrium, as the latter captures full strategic interactions. Finally, numerical simulations on the charging control of plug-in electric vehicles corroborate our theoretical findings.
\end{abstract}


\section{Introduction}      \label{sec:introduction}
Game theory is extensively utilized as a powerful modeling tool to analyze decision-making processes in complex systems where multiple agents interact strategically. The Nash equilibrium (NE), a cornerstone solution concept, is typically referred to as the generalized Nash equilibrium (GNE) in the presence of coupling constraints. In recent years, numerous distributed algorithms relying on local communication have been developed to address GNE seeking problems~\cite{pavel2022distributed}. Representative works include the forward-backward splitting method with fixed step-sizes~\cite{yi2019operator}, the diminishing step-size algorithm over time-varying graphs~\cite{belgioioso2020distributed}, and the tracking-based approach with linear convergence~\cite{carnevale2024tracking}. A comprehensive overview of this field can be found in recent survey papers~\cite{belgioioso2022distributed,hu2022distributed,pavel2025operator}.

The congestion model, originally introduced by Wardrop~\cite{wardrop1952road} for road traffic research, has been extensively studied in the context of transportation systems~\cite{farokhi2015piecewise}, and successfully extended to energy management~\cite{belgioioso2020energy, paccagnan2018nash} and economics~\cite{dafermos1987oligopolistic}. The vanishing individual influence justifies adopting the Wardrop equilibrium (WE), also called the user equilibrium. Unlike the NE, where players internalize their impact on the aggregate, the WE assumes players treat the aggregate as exogenous. This approximation neglects the impact of individual deviations on the aggregate, which simplifies the gradient calculation and thereby significantly improves computational tractability for large-scale problems.

We investigate a class of games known as aggregative congestion games. Let $x_i$ denote the strategy of player $i$, and $\sigma(\boldsymbol{x})$ represent the aggregate strategy of all players. The cost function for the $i$-th player is defined as 
\begin{equation} \nonumber
    J_{i}(x_{i},\sigma(\boldsymbol{x})) = 
    p_{i}(x_{i}) + q\! \left( \sigma(\boldsymbol{x}) \right)^{\top} x_{i}.
\end{equation}
This formulation is termed ``aggregative'' because the coupling between players occurs solely through the aggregate $\sigma(\boldsymbol{x})$, rather than through individual pairwise interactions. Furthermore, the term ``congestion'' arises from the pricing mechanism $q\! \left( \sigma(\boldsymbol{x}) \right)$, which represents a per-unit cost, or congestion price, that typically increases with aggregate demand. The term $p_i$ represents the individual negative utility dependent on $x_i$. This formulation is frequently used in engineering domains~\cite{belgioioso2022distributed}, and a notable example is charging control of electric vehicles (EVs)~\cite{ma2011decentralized}. In this context, $x_i$ represents the individual charging input, while $\sigma(\boldsymbol{x})$ denotes the average EV demand. Consequently, $p_i(x_i)$ serves as a penalty for deviating from a target schedule, and $q\! \left( \sigma(\boldsymbol{x}) \right)$ reflects the energy price.

Moreover, uncertainty is inherent in practical applications. For example, the decision-making process in the aforementioned EV charging control problem is subject to high uncertainties in renewable power generation and charging demand~\cite{jiang2020computing}, as well as electricity price~\cite{fele2020probably}. Since exact solutions are often intractable under uncertainty, robust formulations offer a computationally viable alternative.

To handle these robust formulations, literature on robust equilibrium computation, particularly under uncertain constraints, relies on three distinct paradigms. One builds on scenario optimization~\cite{campi2018introduction}, drawing samples from uncertainty sets to provide solutions with purely probabilistic guarantees~\cite{pantazis2024priori,chen2025inverse}. Another recent direction is distributionally robust optimization~(DRO)~\cite{kuhn2025distributionally}, which bridges the gap by optimizing against a worst-case probability distribution; however, DRO typically leads to severe computational intractability in GNE seeking. For instance, to handle distributionally robust chance constraints, the authors in \cite{fabiani2023distributionally} are forced to formulate a mixed-integer problem, subsequently relying on an artificial dummy agent and delicately tuned penalty functions to relax these binary restrictions back into a continuous setting. In contrast, works such as~\cite{chen2021distributed,fochesato2023generalized,xu2023algorithm} advocate for a worst-case approach by rigorously exploiting the geometric structure of uncertainty sets. This paradigm leverages tools borrowed from robust optimization to inherently preserve a continuous formulation that guarantees deterministic feasibility under all realizations, rather than just probabilistic bounds, without the need for prior data distributions or complex non-convex relaxations.

While the deterministic worst-case paradigm offers superior guarantees, achieving fully distributed equilibrium computation for robust aggregative congestion games remains a critical challenge. To address this, we propose a distributed algorithm to seek the RGWE under structural coupling uncertainties. Since uncertainty hinders the direct computation of standard equilibria, we adopt the RGWE as an inherently tractable alternative, and additionally investigate its quantitative relationship with the robust generalized Nash equilibrium (RGNE).

The main contributions of this paper are as follows.
\begin{enumerate}
    \item To address the uncertainty in coupling constraints, we employ robust optimization theory to derive a tractable deterministic reformulation that accounts for the worst case-realization (Theorem~\ref{thm:robust_trans}). While existing DRO methods (e.g., \cite{fabiani2023distributionally,wen2025distributionally}) often require computationally demanding mixed-integer variables, our formulation uses the augmented complementarity problem to remain continuous and data-free, avoiding computational intractability.
    \item To seek the RGWE, we design a distributed algorithm that integrates a projected primal-dual approach and a dynamic tracking technique (Algorithm~\ref{alg}). This design features an inherent two-time-scale structure, decomposing the system into distinct sub-dynamics.
    \item To analyze the equilibrium obtained by Algorithm~\ref{alg}, we provide theoretical convergence guarantees by leveraging singular perturbation theory and LaSalle's invariance principle (Theorem~\ref{thm:convergence}). Additionally, we characterize the quantitative relationship between RGWE and RGNE (Theorem~\ref{thm:distance}), deriving a distance bound of order $\mathcal{O}(1 {/} \sqrt{N})$.
\end{enumerate}

The remainder of the paper is organized as follows. Section~\ref{sec:preliminaries} introduces the preliminaries, followed by the problem formulation in Section~\ref{sec:WE_with_uncer}. Section~\ref{sec:main_results} presents the main results: Subsection~\ref{sec:exis_and_prop_WE} establishes the deterministic equivalent of the uncertain game; Subsection~\ref{sec:algorithm} develops the distributed algorithm and proves its convergence; and Subsection~\ref{sec:distance} characterizes the approximation bound between RGWE and RGNE. Numerical simulations in Section~\ref{sec:simulation} validate our theoretical results by considering a robust EV charging problem, and finally, Section~\ref{sec:conclusion} concludes the paper.

\section{Preliminaries}      \label{sec:preliminaries}

\subsection{Notations}

$\mathbb{R}^n$ (resp. $\mathbb{R}^{m \times n}$) denotes the set of $n$-dimensional real column vectors (resp. $m \times n$ real matrices).
The positive orthant of $\mathbb{R}^{n}$ is denoted by $\mathbb{R}^{n}_{+}$.
Let $\mathbf{1}_n$ (resp. $\mathbf{0}_n$) denote the all-ones (resp. all-zeros) vector, and let $I_n$ denote the $n \times n$ identity matrix.
For vectors $x_{1},\dots,x_{N}\in \mathbb{R}^{n}$, define $ \operatorname{col}(x_{1}, \dots,x_{N}) :=(x^{\top}_{1}, \dots, x^{\top}_{N})^{\top} \in \mathbb{R}^{Nn}$, 
and let $\| \cdot \|$ denote the Euclidean norm. 
For matrices $A, B \in \mathbb{R}^{m \times m}$, let $A \otimes B$ denote their Kronecker product, while $A \succ B$ ($\succeq B$) indicates that $A - B$ is positive definite (semidefinite). 
Given sets $\mathcal{X}_{1}, \dots, \mathcal{X}_{N} \subseteq \mathbb{R}^{n}$, 
define $\frac{1}{N}\sum_{i=1}^{N} \mathcal{X}_{i} := \{y\in \mathbb{R}^{n}\mid y= \frac{1}{N}\sum_{i=1}^{N}x_{i},\ x_{i} \in \mathcal{X}_{i}\}$.

\vspace{-0.3cm}
\subsection{Convex Analysis}

Let $C \subseteq \mathbb{R}^{n}$ be a closed convex set. For any $x \in \mathbb{R}^{n}$, the Euclidean projection of $x$ onto $C$, denoted by $\Pi_{C}:\mathbb{R}^{n} \to C$, is defined as 
$
    \Pi_{C}(x) := \arg\min_{y\in C} \| x - y \|,
$
which satisfies the nonexpansiveness property
$
    \|  \Pi_{C}(x) - \Pi_{C}(y)  \| \le \|  x-y  \|, \forall x, y \in \mathbb{R}^{n}.
$
For any $x\in C$, the normal cone to $C$ at $x$ is defined by
$
    \mathcal{N}_{C}(x) := \{  v\in \mathbb{R}^{n} \mid v^{\top}(y -x )\le 0,\ \forall y  \in C \}.
$

\begin{prop}    \label{pps:property_of_proj}
    Let $x \in C$ and $s \in \mathbb{R}^{n}$. The following statements are equivalent:
    \begin{enumerate}
        \item $s \in \mathcal{N}_{C}(x)$;
        \item $s^{\top} (y - x) \le 0, \ \forall y \in C  $;
        \item $x = \Pi_{C} (x + s)$.   
    \end{enumerate}  
\end{prop} 
Given $k>0$, a vector-valued function $F:C \to \mathbb{R}^{q}$ is said to be $k$-strongly monotone on $C$ if, for all 
$x, y \in C$,
$$
\left( F(x) - F(y)\right)^{\top}
(x - y) \ge 
k\| x - y \|^2 .
$$
Given $\alpha > 0$, a continuously differentiable function $p: \mathbb{R}^{n} \to \mathbb{R}$ is said to be $\alpha$-strongly convex if, for all $x,y \in \mathbb{R}^{n}$,
$$
    \left(\nabla p(x) - \nabla p(y)\right)^{\top} (x-y) \ge 
    \alpha \|  x -y\|^2.
$$
Given a set $\mathcal{X} \subseteq \mathbb{R}^{p}$ and a mapping $F:\mathcal{X} \to \mathbb{R}^{p}$, the variational inequality problem $\mathrm{VI} (\mathcal{X},F)$ is to find $x \in \mathcal{X}$, such that
$
F(x)^\top(y - x) \ge 0
$
for all $ y \in \mathcal{X}$.

\vspace{-0.3cm}
\subsection{Aggregative Game}   \label{sec:prel:aggregative_games}

Consider a general aggregative game involving $N$ players, indexed by $\mathcal{I} :=  \{1, \dots, N\}$. Each player $i$ chooses its strategy $x_{i}$ subject to a local constraint $\mathcal{X}_{i}$ and a coupling constraint $\mathcal{C}$. Accordingly, the overall constraint for player $i$ is defined as $\Omega_{i}(\boldsymbol{x}_{-i}) := \{x_{i} \in \mathcal{X}_{i} \mid (x_{i},\boldsymbol{x}_{-i}) \in \mathcal{C} \}$, where $\boldsymbol{x}_{-i} = \operatorname{col}(x_{1},\dots,x_{i-1},x_{i+1},\dots,x_{N})$ denotes the collective strategy of all players except $i$.
Player $i$'s cost function is given by $J_i(x_i, \sigma(\boldsymbol{x}))$, where $ \boldsymbol{x} = \operatorname{col}(x_{1},\dots,x_{N})$ is the collective strategy and $\sigma(\boldsymbol{x}) = \tfrac{1}{N} \sum_{j \in \mathcal{I}} x_{j}$ denotes the aggregate of all players' strategies. These definitions lead to the following coupling-constrained game:
\begin{equation}    \label{eq:general_game}
\left\{
\begin{aligned}
    &\text{players:} && \mathcal{I} \\
    &\text{cost function of player } i: && J_i(x_i, \sigma(\boldsymbol{x})) \\
    &\text{overall constraint of player } i: && \Omega_{i}(\boldsymbol{x}_{-i})
\end{aligned}
\right.
\end{equation}
The overall feasible set is $\boldsymbol{\Omega} := \boldsymbol{\mathcal{X}} \cap \mathcal{C} $, where $\boldsymbol{\mathcal{X}} = \prod_{i=1}^{N} \mathcal{X}_i$.

In modeling the behavior of rational players in non-cooperative games, the concept of equilibrium plays a central role. Among the various equilibrium concepts, the $\varepsilon$-generalized Nash equilibrium ($\varepsilon$-GNE) is particularly important.

\begin{definition}[$\varepsilon$-GNE]     \label{def:GNE}
    Given $\varepsilon \ge 0$, a collective strategy $\boldsymbol{x}^{\rm N} \in \boldsymbol{\Omega}$ is an $\varepsilon$-GNE of the aggregative game~\eqref{eq:general_game} if, for all $i \in \mathcal{I}$ and all $x_{i} \in \Omega_{i}(\boldsymbol{x}^{\rm N}_{-i})$:
    \begin{equation}         \nonumber
        J_{i}(x^{\rm N}_{i},\sigma(\boldsymbol{x}^{\rm N})) \leq J_{i}\left(x_{i},\frac{1}{N}x_{i}+\frac{1}{N}\sum_{j=1,j\neq i}^{N}x_{j}^{\rm N}\right) + \varepsilon.
    \end{equation}
Specifically, $\boldsymbol{x}^{\rm N}$ is said to be a GNE if $\varepsilon = 0$.       
\end{definition}

Roughly speaking, GNE (resp. $\varepsilon$-GNE) represents a stable solution from which no player can unilaterally change their strategy to reduce their cost (resp. to reduce their cost by at least $\varepsilon$). 

As the number of players increases, the influence of each individual player becomes negligible over the network. Consequently, each player responds solely to the aggregate, rather than to the strategies of individual players. This behavior gives rise to the concept of Wardrop equilibrium (WE). When extended to include coupling constraints, this concept becomes the generalized Wardrop equilibrium (GWE)~\cite{paccagnan2018nash,bakhshayesh2023generalized}.

\begin{definition}[GWE]   \label{def:GWE}
A collective strategy $\boldsymbol{x}^{\rm W} \in \boldsymbol{\Omega}$ is a GWE of the aggregative game \eqref{eq:general_game} if, for all $i \in \mathcal{I}$ and all $x_i \in \Omega_i(\boldsymbol{x}^{\rm W}_{-i})$:
\begin{equation} \nonumber
J_i(x^{\rm W}_i, \sigma(\boldsymbol{x}^{\rm W})) \leq J_i(x_i, \sigma(\boldsymbol{x}^{\rm W})).    
\end{equation}        
\end{definition}  

GWE distinguishes itself from GNE by its lower computational complexity, as players evaluating deviations need not account for the coupled effects of others' strategies. This renders GWE particularly suitable for large-scale, nonatomic systems~\cite{jacquot2020efficient}.

\vspace{-0.3cm}
\subsection{Graph Theory}

Consider an undirected graph $G = (\mathcal{V},\mathcal{E})$, where $\mathcal{V} = \{ 1,\dots,N \}$ denotes the set of nodes and $\mathcal{E}\subseteq \mathcal{V}\times \mathcal{V}$ represents the set of edges. 
The graph $G$ is associated with a non-negative weighted adjacency matrix $W=[w_{ij}]\in \mathbb{R}^{N \times N}$, such that $w_{ij} = w_{ji} > 0$ if $\{i,j\}\in \mathcal{E}$ and $w_{ij} = 0$ otherwise. 
The matrix $W$ is called doubly stochastic if it satisfies
$
    W \mathbf{1}_{N} =  \mathbf{1}_{N} $ and $ \mathbf{1}_{N}^{\top} W =  \mathbf{1}_{N}^{\top}.
$
The Laplacian matrix is defined as $L=\Delta - W$, where $\Delta=\operatorname{diag}(\delta_{1},\dots,\delta_{N} )\in \mathbb{R}^{N \times N}$ is a diagonal matrix, with the $i$-th diagonal element given by $\delta_{i}=\sum^{N}_{j=1}w_{ij}$. 
Moreover, if the graph $G$ is connected, i.e., there exists a path between any two nodes, then the Laplacian matrix satisfies $L=L^{\top} \succeq 0$, with a single zero eigenvalue and all other eigenvalues positive. In particular, when $W$ is doubly stochastic, we have $\Delta = I$, yielding
$
    L \mathbf{1}_{N} = \mathbf{0}_{N} $ and $ \mathbf{1}_{N}^{\top} L = \mathbf{0}_{N}^{\top}.
$

\section{Problem Formulation}  \label{sec:WE_with_uncer}
In this section, we formulate an aggregative congestion game subject to uncertain coupling constraints.

Consider an aggregative congestion game $\mathcal{G}$ of $N$ players. Each player $i \in \mathcal{I}$ selects a strategy $x_{i}$ from its local constraint set $\mathcal{X}_{i} \subseteq \mathbb{R}^{n}$, while $\boldsymbol{x} \in \boldsymbol{\mathcal{X}}$ denotes the collective strategy. 
Player $i$'s cost function $J_{i}:\mathbb{R}^{n}\times \mathbb{R}^{n} \to \mathbb{R}$ is continuously differentiable and given by
\begin{equation} \label{eq:cost_function}
    J_{i}(x_{i},\sigma(\boldsymbol{x})) = 
    p_{i}(x_{i}) + q\! \left( \sigma(\boldsymbol{x}) \right)^{\top} x_{i},
\end{equation}
where 
$p_{i}:\mathcal{X}_{i} \to \mathbb{R}$ and $q:\tfrac{1}{N}\sum^{N}_{j=1} \mathcal{X}_{j} \to \mathbb{R}^{n}$ 
denote, respectively, the negative utility associated with $x_{i}$ and per-unit cost as a function of the aggregate $\sigma(\boldsymbol{x})$. 

Given $\boldsymbol{x}_{-i}$, player $i$'s strategy $x_{i}$ is subject to the coupling constraint
\begin{equation} \nonumber
    \mathcal{C}_{i,\boldsymbol{\rho}}(\boldsymbol{x}_{-i})
    :=\left\{\,x_i\in\mathcal{X}_i\ \middle|\ \rho_i^{\top}x_i \le
    b - \sum_{j=1,j\neq i}^{N}\rho_j^{\top}x_j \,\right\},
\end{equation}
where the parameter vector is $\boldsymbol{\rho} = \operatorname{col}(\rho_{1},\dots,\rho_{N}) \in \mathbb{R}^{Nn}$, and $b$ is a scalar. 
Accordingly, the overall coupling constraint can be written as 
\begin{equation}            \label{eq:coupling_constraints_overall}
    \mathcal{C}_{\boldsymbol{\rho}} 
    = \left\{\boldsymbol{x} \in \boldsymbol{\mathcal{X}}
    \ \middle| \  \sum_{i=1}^{N}\rho_{i}^\top x_{i}\le b \right\}.
\end{equation}
This affine coupling constraint~\eqref{eq:coupling_constraints_overall} is widely adopted in the literature (e.g., \cite{yi2019operator,belgioioso2020distributed,carnevale2024tracking}) and covers applications involving bounded shared resources (e.g., \cite{paccagnan2018nash,ma2011decentralized}).
%
%
%
%
%

\begin{comment}
    
Thus,  given $\boldsymbol{x}_{-i}$, the $i$-th player solves the following problem $\mathcal{G}$
\begin{equation}    \label{eq:uncertain_game}
    \begin{aligned}
        & \min_{x_{i}} \quad p_{i}(x_{i}) + q\left( \sigma(\boldsymbol{x}) \right)^\top x_{i}    \\
        %
        %
        & \begin{array}{r@{\quad}r@{}l@{\quad}l}
        %
        \text{s.t.} & x_{i}\in\mathcal{C}_{i,\boldsymbol{\rho}}(\boldsymbol{x}_{-i}),  
        %
        \ \boldsymbol{\rho} \in \boldsymbol{P}.
        \end{array}
    \end{aligned}    
\end{equation}

\end{comment}

Given the parameter $\boldsymbol{\rho}$, the aggregative congestion game $\mathcal{G}$ can be described as
\begin{equation}    \label{eq:uncertain_game}
\mathcal{G} \colon
\left\{
\begin{aligned}
    &\text{players:} && \mathcal{I} \\
    &\text{cost function of player } i: && p_{i}(x_{i}) + q\! \left( \sigma(\boldsymbol{x}) \right)^\top x_{i} \\
    &\text{overall constraint of player } i : && \mathcal{C}_{i,\boldsymbol{\rho}}(\boldsymbol{x}_{-i})
\end{aligned}
\right.
\end{equation}

It is important to note that each player's strategy $x_{i}$ directly affects the aggregate, and, consequently, influences the cost functions of all players.
Nevertheless, in aggregative congestion games, the influence of an individual player's strategy decreases as the population size $N$ grows \cite{devarajan1981note}. In this limit, the game $\mathcal{G}$ in \eqref{eq:uncertain_game} comes to a nonatomic model, in which players are infinitesimal agents forming a continuum, thereby allowing significant simplifications in solution concepts~\cite{jacquot2020efficient}.
Meanwhile, the growth of $N$ also introduces practical challenges: the computational burden of solving GNE problems scales rapidly with $N$ \cite{paccagnan2018nash}, motivating the need for scalable and tractable equilibrium concepts.
Consequently, GWE in Definition~\ref{def:GWE} serves as a natural solution that balances accuracy and computational efficiency.
%
%It captures limiting behavior in large populations while ensuring a close approximation to the GNE.
%
%
%

Now, we further consider $\boldsymbol{\rho}$ as an environmental parameter affected by uncertainty, which renders the aggregative congestion game $\mathcal{G}$ uncertain. Specifically, the uncertain feasibility set
$\boldsymbol{P} = \prod_{i=1}^{N}P_{i}$,
where each $P_{i} \subseteq \mathbb{R}^{n}$ is a polyhedral set characterizing the possible values of $\rho_{i}$, i.e.,
\begin{equation}   \nonumber
    P_{i} =\{ \rho _{i} \in \mathbb{R}^{n} \mid D_{i} \rho_{i} \le d_{i} \},
\end{equation}
where  the rows of $D_{i}\in \mathbb{R}^{m\times n}$ are normalized, and $d_{i}\in \mathbb{R}^{m}$.

To address this uncertainty, we introduce the concept of robust generalized Wardrop equilibrium (RGWE), defined below, which ensures robustness against worst-case parameter realizations.

\begin{definition}[RGWE]   \label{def:RWE}
    A collective strategy 
    $\boldsymbol{x}^{\rm W} \in \bigcap_{ \boldsymbol{\rho} \in \boldsymbol{P}}\mathcal{C}_{\boldsymbol{\rho}} $
    is an RGWE of the uncertain aggregative congestion game $\mathcal{G}$
    if, 
    for all $i\in \mathcal{I}$ and all $x_i \in \bigcap_{\boldsymbol{\rho} \in \boldsymbol{P}} \mathcal{C}_{i,\boldsymbol{\rho}}(\boldsymbol{x}_{-i}^{\rm W})$:
    \begin{equation}    \nonumber
    \begin{aligned}
        p_{i}(x^{\rm W}_{i}) + q\! \left( \sigma(\boldsymbol{x}^{\rm W}) \right)^\top   \!x_{i}^{\rm W}
        \! \leq 
        p_{i}(x_{i}) & + q\! \left( \sigma(\boldsymbol{x}^{\rm W}) \right)^\top \! x_{i}.  
    \end{aligned}     
    \end{equation}        
\end{definition}

In this paper, our main objective is to develop a distributed algorithm to compute the RGWE for the aggregative congestion game $\mathcal{G}$ in \eqref{eq:uncertain_game} subject to constraint uncertainties, and to provide its convergence guarantee and equilibrium analysis. The following assumptions form the basis of our study.
\begin{assumption}    \label{asp:convexity_and_nonempty}
    \begin{enumerate}
    \item For $i \in \mathcal{I}$, the set $\mathcal{X}_{i}$ is nonempty, compact, and convex.
    \item For $i \in \mathcal{I}$, 
    $p_{i}(\cdot)$ is $\alpha$-strongly convex over $\mathcal{X}_{i}$, while $q(\cdot)$ is $\beta$-strongly monotone and $L_{q}$-Lipschitz continuous over $\mathbb{R}^{n}$.
    \item The intersection $\bigcap_{ \boldsymbol{\rho} \in \boldsymbol{P}}\mathcal{C}_{\boldsymbol{\rho}}$ has a nonempty interior point.
    \item The undirected graph $G$ is connected, and the weighted adjacency matrix $W$ is doubly stochastic.       
    \end{enumerate}
\end{assumption}

Specifically, Assumption 1.1) imposes requirements on the local constraints, a common practice within the game-theoretic literature~\cite{jacquot2020efficient,paccagnan2018nash,pantazis2024priori,xu2022efficient,belgioioso2020distributed,koshal2016distributed}.
Furthermore, Assumption 1.2) is consistent with existing studies on the cost function~\eqref{eq:cost_function}, such as~\cite{paccagnan2018nash,ma2011decentralized,bakhshayesh2023generalized}.
Additionally, Assumption 1.3), known as Slater's condition, is considered mild in robust game settings~ \cite{pantazis2024priori,chen2025inverse,chen2021distributed}. Since it holds for the entire uncertain set $\boldsymbol{P}$, it naturally holds given any fixed $\boldsymbol{\rho}$.
Lastly, Assumption 1.4) is widely used in distributed NE seeking algorithms for aggregative games~\cite{carnevale2024tracking,belgioioso2020distributed,koshal2016distributed}.

\section{Main Results}    \label{sec:main_results}
In this section, we present the main results of this paper. Subsection~\ref{sec:exis_and_prop_WE} casts the uncertain constraint \eqref{eq:coupling_constraints_overall} as an augmented deterministic formulation. Subsection~\ref{sec:algorithm} proposes a distributed algorithm with convergence guarantees under fixed step-sizes. Finally, Subsection~\ref{sec:distance} quantifies the relationship between RGWE and RGNE w.r.t. the population size.

\subsection{Augmented Reformulation} \label{sec:exis_and_prop_WE}
Analyzing the uncertain game $\mathcal{G}$ is non-trivial due to the requirement of feasibility for all uncertainty realizations.
To address this, we employ tools from robust optimization~\cite{bertsimas2011theory} to transform $\mathcal{G}$ into a deterministic and tractable game. This reformulation builds on the following theorem, with detailed proof in Appendix~\ref{appen:A}.

\begin{theorem}    \label{thm:robust_trans}
    Under Assumption \ref{asp:convexity_and_nonempty}, a collective strategy $\boldsymbol{x}$ is feasible to the coupling constraint \eqref{eq:coupling_constraints_overall} under all possible uncertainties if and only if there exists $\boldsymbol{\gamma} = \operatorname{col}(\gamma _{1}, \dots, \gamma _{N}) \in \mathbb{R}^{Nm}_{+}$ such that, $(\boldsymbol{x} , \boldsymbol{\gamma})$ is feasible to the following augmented constraints
    \begin{equation}     \label{eq:augmented_game}
        \begin{aligned} 
            \ &  \sum_{j=1}^{N}d_{j}^\top \gamma_{j} \le b, \ D_{i}^\top\gamma_{i} - x_{i}=\mathbf{0}_{n},\ \forall i \in \mathcal{I}.
        \end{aligned}    
    \end{equation}
\end{theorem}

Leveraging Theorem~\ref{thm:robust_trans}, we can now reformulate the original uncertain game $\mathcal{G}$. For ease of notation, we combine the decision variables $\boldsymbol{x}$ and $\boldsymbol{\gamma}$ from condition (\ref{eq:augmented_game}) into an augmented vector 
$\boldsymbol{y} = \operatorname{col}(y_{1},\dots,y_{N}) $, where $y_{i} = \operatorname{col}(x_{i},\gamma _{i}) \in \mathcal{Y}_{i}$ with $\mathcal{Y}_{i} = \mathcal{X}_{i} \times \mathbb{R}^{m}_{+}$.
Condition (\ref{eq:augmented_game}) defines the augmented constraint set:
\begin{equation}   \nonumber
    \Omega _{i}(\boldsymbol{y}_{-i}) : = \left\{y_{i} \in \mathcal{Y}_{i} %
    \ \middle| \ \sum_{j=1}^{N} A_{j}^{\top}y_{j} \leq b, 
      B_{i}^{\top}y_{i} = \mathbf{0}_{n} \right\},    \forall i \in \mathcal{I}.
\end{equation}
Accordingly, we define the vector $A_{i} = 
\operatorname{col} (\mathbf{0}_{n}, d_{i}) \in \mathbb{R}^{n+m}$
and
matrix $B_{i} = 
\begin{bmatrix}
    -I_{n} \\ D_{i}
\end{bmatrix} \in \mathbb{R}^{(n+m)\times n}.
$
Consequently, with
$\boldsymbol{\mathcal{Y}} = \prod_{i=1}^{N} \mathcal{Y}_{i} $, 
the overall constraint set is given by
\begin{equation}   \nonumber
    \boldsymbol{\Omega} = \left\{ \boldsymbol{y} \in  
    \boldsymbol{\mathcal{Y}} \ \middle| \ \sum_{j=1}^{N} A_{j}^{\top}y_{j} \leq b, \  
    B_{i}^{\top}y_{i} = \mathbf{0}_{n} , \ \forall i \in \mathcal{I} \right\}.
\end{equation}
Moreover, we define the augmented cost functions $\bar{p}_i(y_i)$, $\bar{q}(\bar{\sigma}(\boldsymbol{y}))$ and the corresponding aggregate mapping $\bar{\sigma}(\boldsymbol{y})$ as:
\begin{enumerate}
    \item For each $i \in \mathcal{I}$, the local cost $\bar{p}_{i}: \mathcal{Y}_{i} \to \mathbb{R}$ is defined as $\bar{p}_{i}(y_{i}) = p_{i}(x_{i})$.
    \item The aggregate in the augmented space is defined as $\bar{\sigma}(\boldsymbol{y}) : = \frac{1}{N}\sum_{i=1}^{N}y_{i} = \operatorname{col} \left( \sigma(\boldsymbol{x}), \frac{1}{N}\sum_{i=1}^{N} \gamma_{i} \right)$.
    \item The coupling cost is reformulated via the augmented function $\bar{q}(\bar{\sigma} (\boldsymbol{y}) ) = \operatorname{col}( q(\sigma (\boldsymbol{x})), \mathbf{0}_m )$.
\end{enumerate}

It follows from Definition \ref{def:RWE} and Theorem \ref{thm:robust_trans} that the problem of computing an RGWE of $\mathcal{G}$ is equivalent to finding a GWE in Definition \ref{def:GWE} of the following augmented game 
\begin{equation}   \label{eq:transfer_final_game}
\bar{\mathcal{G}} \colon
\left\{
\begin{aligned}
    &\text{players:} && \mathcal{I} \\
    &\text{cost function of player } i: && \bar{p}_{i}(y_{i}) + \bar{q} \! \left( \bar{\sigma}(\boldsymbol{y}) \right)^\top y_{i} \\
    &\text{overall constraint of player }i: && \Omega _{i}(\boldsymbol{y}_{-i})
\end{aligned}
\right.
\end{equation}

We now define the operator $ \bar{\boldsymbol{F}}^{\rm W} : \boldsymbol{\mathcal{Y}} \to \mathbb{R}^{N(n + m)}$ associated with the GWE of the augmented game $\bar{\mathcal{G}}$. This operator is constructed by stacking the partial gradients of each player’s cost function, evaluated under a fixed aggregate $\sigma(\boldsymbol{y})$:
\begin{equation}  \label{eq:stacked_gradient}
    \bar{\boldsymbol{F}}^{\rm W}(\boldsymbol{y}) 
    = \operatorname{col}\left(\bar{F}^{\rm W}_{1}(\boldsymbol{y}),  \dots,  \bar{F}^{\rm W}_{N}(\boldsymbol{y})\right),
\end{equation}
where for all $i \in \mathcal{I}$,
\begin{equation}     \label{eq:i_gradient_Wardrop}
    \bar{F}^{\rm W}_{i}(\boldsymbol{y}) = 
    \begin{bmatrix}  
        \nabla p_{i}(x_{i}) + q( \sigma(\boldsymbol{x}))\\ 
        \mathbf{0}_{m}
    \end{bmatrix}\in \mathbb{R}^{n+m}.
\end{equation}
Given the operator \eqref{eq:stacked_gradient}, Assumption \ref{asp:convexity_and_nonempty} ensures that the conditions of \cite[Prop. 1]{paccagnan2018nash} are met. Furthermore, invoking \cite[Prop. 3.2.1]{facchinei2007finite}, Slater's condition guarantees the boundedness of the Lagrange multipliers $\boldsymbol{\gamma}$, which implies the compactness of $\boldsymbol{\Omega}$. Consequently, \cite[Cor. 2.2.5]{facchinei2007finite} ensures a nonempty and compact solution set to \(\mathrm{VI}(\boldsymbol{\Omega},\bar{\boldsymbol{F}}^{\rm W})\). Thus, \cite[Prop. 1]{paccagnan2018nash} establishes that any solution \(\boldsymbol{y}^{\rm W}\) to \(\mathrm{VI}(\boldsymbol{\Omega},\bar{\boldsymbol{F}}^{\rm W})\) is a variational GWE of \(\bar{\mathcal{G}}\), which, by Theorem~\ref{thm:robust_trans}, induces an RGWE of \(\mathcal{G}\). We formalize this result in the following proposition:

\begin{prop}  \label{prop:exi_of_WE}
    Under Assumption \ref{asp:convexity_and_nonempty}, a solution \(\boldsymbol{y}^{\rm W}\) to \(\mathrm{VI}(\boldsymbol{\Omega},\bar{\boldsymbol{F}}^{\rm W})\) induces an RGWE of the uncertain game \(\mathcal{G}\).
\end{prop}

\begin{comment}
    \begin{IEEEproof}
    All the assumptions of \cite[Prop. 1]{paccagnan2018nash} are satisfied by Assumption~\ref{asp:convexity_and_nonempty}, thus any solution \(\boldsymbol{y}^{\rm W}\) of \(\mathrm{VI}(\boldsymbol{\Omega},\bar{\boldsymbol{F}}^{\rm W})\) is a variational GWE of the augmented game \(\bar{\mathcal{G}}\). By Theorem~\ref{thm:robust_trans}, this solution \(\boldsymbol{y}^{\rm W}\) then induces an RGWE \(\boldsymbol{x}^{\rm W}\) of \(\mathcal{G}\).
\end{IEEEproof}
\end{comment}
%
%
%
%
%
%
%

It follows by \cite[Th. 3.1]{facchinei2007generalized} that, a collective strategy $\boldsymbol{y}$
    is a variational GWE of the augmented game $\bar{\mathcal{G}}$,
    if and only if there exists $\lambda_{i} \in \mathbb{R}^{n}$ and $\mu_{i}\in \mathbb{R}_{+}$ such that, for all $i\in \mathcal{I}$,
    \begin{equation}           \nonumber
        \begin{aligned}
            & \mathbf{0}_{n + m} \in \bar{F}^{\rm W}_{i}(\boldsymbol{y}) 
            + A_{i}\mu_{i} + B_{i}\lambda_{i} + 
            \mathcal{N}_{\mathcal{Y}_{i}}(y_{i}), \\
            & 0 \ge \sum_{j=1}^{N} A_{j}^{\top}y_{j} - b\perp \mu _{i}, \\
            & \mathbf{0}_{n} = B^{\top}_{i}y_{i},
        \end{aligned}
    \end{equation}
where $\mu_{1} =  \mu_{2} = \dots = \mu_{N}$. For connected graphs, $L\boldsymbol{\mu} = \mathbf{0}_{N}$ enforces consensus, yielding the compact Karush--Kuhn--Tucker~(KKT) conditions:
\begin{subequations}   \label{eq:KKT}
    \begin{align}
        & \mathbf{0}_{N( n+ m )} \in \bar{\boldsymbol{F}}^{\rm W}(\boldsymbol{y}) 
        + \boldsymbol{A}\boldsymbol{\mu} 
        + \boldsymbol{B}\boldsymbol{\lambda}
        +\mathcal{N}_{\boldsymbol{\mathcal{Y}}}(\boldsymbol{y}), \label{eq:KKT:a}\\
        & 0 \ge \mathbf{1}_{N}^{\top} 
        (\boldsymbol{A}^{\top}\boldsymbol{y} - \boldsymbol{b}),
        \quad 0 = (\boldsymbol{A}^{\top}\boldsymbol{y} - \boldsymbol{b})^{\top}\boldsymbol{\mu} , \\
        & \mathbf{0}_{Nn} = \boldsymbol{B}^{\top}\boldsymbol{y}, \\
        & \mathbf{0}_{N} = L\boldsymbol{\mu}, \label{eq:KKT:d}
    \end{align}
\end{subequations}
where 
$\boldsymbol{A} = \operatorname{diag}(A_{1},\dots,A_{N}) \in \mathbb{R}^{N(n + m) \times N},\ 
\boldsymbol{B} = \operatorname{diag}(B_{1},\dots,B_{N}) \in \mathbb{R}^{ N(n + m) \times Nn},\ 
\boldsymbol{b}=\operatorname{col}(b_{1},\dots,b_{N})\in \mathbb{R}^{N}  $,
$b=\sum^{N}_{j=1}b_{j} $, 
$\boldsymbol{\mu} = \operatorname{col}(\mu_{1},\dots,\mu_{N})\in \mathbb{R}^{N}_{+} $, 
and $\boldsymbol{\lambda} = \operatorname{col}(\lambda_{1},\dots,\lambda_{N})\in \mathbb{R}^{Nn} $. 
%
%
%
%
%

%\vspace*{15\baselineskip} 
\subsection{Distributed Algorithm}      \label{sec:algorithm}
Theorem \ref{thm:robust_trans} and Proposition \ref{prop:exi_of_WE} establish the equivalence between RGWE of $\mathcal{G}$ and variational GWE of $\bar{\mathcal{G}}$. Leveraging this, we can design a distributed algorithm to seek RGWE of $\bar{\mathcal{G}}$ by solving the KKT conditions~\eqref{eq:KKT}.

A key challenge in addressing the aggregative congestion game~$\mathcal{G}$ in a distributed manner is that players lack direct access to the true aggregate $\bar{\sigma}(\boldsymbol{y})$, necessitating its estimation via partial-decision information. Inspired by dynamic tracking techniques~\cite{carnevale2024tracking,belgioioso2020distributed,koshal2016distributed}, each player maintains an auxiliary variable $\hat{\sigma}^{k}_{i} $ to estimate the aggregate $\bar{\sigma}(\boldsymbol{y})$ at time $k$. This variable evolves jointly with the player's own strategy and is updated through local communication according to
\begin{equation}  \nonumber
    \hat{\sigma}^{k+1}_{i} =  \sum\nolimits_{j \in \mathcal{I}}w_{ij}\hat{\sigma}^{k}_{j} 
            + y_{i}^{k + 1} - y_{i}^{k}.
\end{equation}
For notational convenience, we define $\phi_{i}(y_{i},\bar{\sigma}(\boldsymbol{y})) := \bar{F}^{\rm W}_{i}(\boldsymbol{y})$ to make the dependence on the aggregate explicit. Then, player $i$ evaluates  \(\phi_i(y_i^k,\hat\sigma_i^k)\) by substituting the true aggregate with the estimate $\hat{\sigma}^{k}_{i}$.

In this paper, each player updates their decision via a modified projected primal-dual method. Specifically, this method endows each player $i$ with a local copy of the concentrated multiplier $\mu_{i} $, and employs a local auxiliary variable $\nu_{i}$ to reach consensus on $\mu_{i}$. The algorithm is designed as follows:
\begin{algorithm}   [htbp]
	%\renewcommand{\thealgorithm}{1} 
	%\SetAlgoRefName{} % no count number
	\caption{Distributed RGWE Seeking Algorithm}
	\label{alg}
	\vspace{0.1cm}

    \begin{minipage}{\linewidth}
    \textbf{Initialization}: $k \gets 0$. For all $i \in \mathcal{I}$, set $y^{0}_{i}  \in \mathcal{Y}_{i},\
    \nu^{0}_{i}  \in \mathbb{R},\
    \mu^{0}_{i} \in \mathbb{R}_{+},\
    \lambda ^{0}_{i} \in \mathbb{R}^{n}, $
    and $\hat{\sigma}^{0}_{i} = y^{0}_{i} \in \mathbb{R}^{n+m}.$ Fixed step-size $\delta$.
    \end{minipage}

    \vspace{0.15cm}
    
    \OuterLoop{
        \primaldual{  \vspace{-0.25cm}
            \begin{flalign*}
                & y^{k+1}_{i} =  
                \Pi_{\mathcal{Y}_{i}} 
                \left[y^{k}_{i} - \delta ( \phi_{i}(y^{k}_{i},\hat{\sigma}^{k}_{i}) + A_{i}\mu^{k}_{i} + B_{i}\lambda^{k}_{i})\right] , & \\
                & \nu_{i}^{k + 1}  =  \nu_{i}^{k} + \delta \sum_{j \in \mathcal{I}}w_{ij} 
                (\mu ^{k}_{i} - \mu ^{k}_{j}  ), & \\
                & \mu ^{k+1}_{i}  =  \Pi_{\mathbb{R}_{+}} \Big[ \mu ^{k}_{i} 
                + \delta (A_{i}^{\top} y_{i}^{k} - b_{i}  )  \\
                & \quad \quad \quad - \delta \sum_{j \in \mathcal{I}}w_{ij}
                (\mu ^{k}_{i} - \mu ^{k}_{j}  )
                - \delta \sum_{j \in \mathcal{I}}w_{ij} 
                \left( \nu_{i}^{k} - \nu_{j}^{k}  \right)  \Big], & \\
                & \lambda ^{k+1}_{i} =   \lambda^{k}_{i} 
                + \delta  B^{\top}_{i} y^{k}_{i}, &
            \end{flalign*}
            \vspace{-0.5cm}
            }

        \vspace{-0.05cm}
        
        \dynamictracking{\vspace{-0.15cm}
            \begin{flalign*}
                & \hat{\sigma}^{k+1}_{i} = \sum_{j \in \mathcal{I}}w_{ij}\hat{\sigma}^{k}_{j} 
                + y_{i}^{k + 1} - y_{i}^{k}, &
            \end{flalign*}
            \vspace{-0.15cm}
            }
        \vspace{-0.15cm}
        
        $k \leftarrow k + 1$\;
    }
\end{algorithm}
\vspace{-0.35cm}

From an entire aspect, Algorithm~\ref{alg} can also be expressed more clearly in a compact form:
\begin{subequations}        \label{eq:algorithm_compact}
    \begin{align}
        &\boldsymbol{y}^{k+1} = 
        \Pi_{\boldsymbol{\mathcal{Y}}} \left[ \boldsymbol{y}^{k}
        - \delta \left( \boldsymbol{\phi}(\boldsymbol{y}^{k}, \hat{\boldsymbol{\sigma}}^{k} ) + \boldsymbol{A} \boldsymbol{\mu}^{k} + \boldsymbol{B}\boldsymbol{\lambda}^{k} \right) \right] ,   \label{eq:alg:a} \\
        & \boldsymbol{\nu}^{k + 1} = \boldsymbol{\nu}^{k}  + \delta L \boldsymbol{\mu}^{k}, \\
        & \boldsymbol{\mu}^{k + 1} = \Pi_{\mathbb{R}_{+}^{N}}  
        \left[ \boldsymbol{\mu}^{k} 
        + \delta (\boldsymbol{A}^{\top} \boldsymbol{y}^{k} - \boldsymbol{b}  ) 
        - \delta L \boldsymbol{\mu}^{k} - \delta L \boldsymbol{\nu}^{k} \right], \\
        & \boldsymbol{\lambda}^{k + 1} = \boldsymbol{\lambda}^{k} 
        + \delta  \boldsymbol{B}^{\top} \boldsymbol{y}^{k},   \label{eq:alg:d} \\
        & \hat{\boldsymbol{\sigma}}^{k+1} = (W \otimes I_{n+m}) \hat{\boldsymbol{\sigma}}^{k} 
        +  \boldsymbol{y}^{k+1} - \boldsymbol{y}^{k},   \label{eq:alg:e}
    \end{align}
\end{subequations}
with the same compact notations as in (\ref{eq:KKT}), where
$\boldsymbol{\phi}(\boldsymbol{y}^{k}, \hat{\boldsymbol{\sigma}}^{k} ) 
= \operatorname{col} \left( \phi_{1}(y_{1}^{k}, \hat{\sigma}^{k}_{1}),\ \dots, \ \phi_{N}(y_{N}^{k}, \hat{\sigma}^{k}_{N})  \right)$ and 
$ \hat{\boldsymbol{\sigma}}^{k} = \operatorname{col} \left(\hat{\sigma}^{k}_{1}, \dots, \hat{\sigma}^{k}_{N} \right)$.

The proposed distributed discrete-time Algorithm~\ref{alg} can be analyzed through the lens of singular perturbation theory, where the fixed step-size $\delta$ acts as the constant perturbation parameter that induces a time-scale separation. Within this framework, \eqref{eq:algorithm_compact} can be interpreted as slow dynamics (\ref{eq:alg:a})-(\ref{eq:alg:d}) interconnected with fast dynamics (\ref{eq:alg:e}). Specifically, the slow dynamics implement the modified projected primal-dual method with consensus, while the fast dynamics ensure aggregate tracking.

\begin{remark}   \label{rmk:alg_compare}
    Notably, \cite{carnevale2024tracking} also employs a singular perturbation approach to design a projection-free primal-dual method based on a piecewise augmented Lagrangian. To ensure dual non-negativity, their algorithm requires the condition $w_{ii} > \delta$. In contrast, our projected primal-dual method inherently enforces non-negativity via the projection operator $\Pi_{\mathbb{R}_{+}^{N}}$. Furthermore, whereas \cite{carnevale2024tracking} relies on a full-rank assumption on matrix $\boldsymbol{A}$ for convergence, we relax this requirement. By invoking  LaSalle's invariance principle, we show that Algorithm~\ref{alg} accommodates problems with linearly-dependent coupling constraints. However, the cost of this relaxation is the loss of a guaranteed linear convergence rate established in \cite{carnevale2024tracking}.
\end{remark}

The convergence guarantee of Algorithm~\ref{alg} is formalized below, with the proof provided in Appendix~\ref{appen:B}.
\begin{theorem}    \label{thm:convergence}
    Under Assumption \ref{asp:convexity_and_nonempty}, there exists a constant step-size $\delta > 0$, such that the sequence $\{ \boldsymbol{y}^{k} \}$ generated by Algorithm \ref{alg} converges to an RGWE of the uncertain aggregative congestion game $\mathcal{G}$.
\end{theorem}
%
%
%

\begin{comment}
\begin{remark}
    One may naturally question whether Algorithm \ref{alg} remains valid for cost functions different from \eqref{eq:cost_function}, since the proof in Appendix \ref{appen:B} leverages the structure of \eqref{eq:cost_function} to derive \eqref{eq:pseudo_strongly_monotone}. In fact, the gradient in \eqref{eq:i_gradient_Wardrop} provides only a sufficient condition to guarantee convergence of Algorithm \ref{alg}. More generally, any cost function with a strongly monotone gradient yields a result analogous to \eqref{eq:pseudo_strongly_monotone}, thereby ensuring convergence for both NE and WE. Furthermore, this proof, based on singular perturbation, does not require compactness of the local constraint sets and guarantees convergence with a fixed step-size.
\end{remark}
\end{comment}

\subsection{RGWE vs. \texorpdfstring{$\varepsilon$-RGNE}{ε-RGNE}}    \label{sec:distance}
This subsection investigates the quantitative relationship between RGWE and RGNE in the uncertain aggregative congestion game~$\mathcal{G}$. Given its role as a fundamental stability concept, the RGNE serves as a standard benchmark for evaluating other equilibrium concepts. However, seeking an RGNE can be computationally intensive. Consequently, as discussed in Sections \ref{sec:prel:aggregative_games} and \ref{sec:WE_with_uncer}, the RGWE is often adopted as a more tractable alternative. The lower computational complexity of the RGWE stems from its formulation, wherein each player considers only unilateral changes to their own strategy. 

To see this more clearly, let us examine the structure of the pseudo-gradient associated with the NE by
\begin{equation}     \nonumber
    \bar{\boldsymbol{F}}^{\rm N}(\boldsymbol{y}) 
    = \operatorname{col}\left(\bar{F}^{\rm N}_{1}(\boldsymbol{y}), \dots , \bar{F}^{\rm N}_{N}(\boldsymbol{y})\right),
\end{equation}
where for all $i \in \mathcal{I}$,
\begin{equation}    \label{eq:i_gradient_Nash}
\bar{F}^{\rm N}_{i} (\boldsymbol{y}) =
    \begin{bmatrix}
        \nabla p_{i}(x_{i}) + 
        \frac{1}{N}\nabla_{z}q(z)_{|z=\sigma(\boldsymbol{x})} x_{i} + 
        q( \sigma(\boldsymbol{x}))   \\  
        \mathbf{0}_{m}
    \end{bmatrix}.
\end{equation}

\begin{prop}[\texorpdfstring{\cite[Th. 2.1]{facchinei2007generalized}}{}]     \label{prop:exi_of_NE}
Under Assumption \ref{asp:convexity_and_nonempty}, any solution \(\boldsymbol{y}^{\rm N}\) of \(\mathrm{VI}(\boldsymbol{\Omega},\bar{\boldsymbol{F}}^{\rm N})\) induces an RGNE of \(\mathcal{G}\).   
\end{prop}

For each player $i$, the computational complexity of $\bar{F}_{i}^{\rm W}$ in \eqref{eq:i_gradient_Wardrop} involves approximately three dominant $\mathcal{O}(n)$ operations, compared to five for $\bar{F}_{i}^{\rm N}$ in \eqref{eq:i_gradient_Nash}. Consequently, a single evaluation of $\bar{F}_{i}^{\rm N}$ requires $67\%$ more computational overhead than $\bar{F}_{i}^{\rm W}$. This computational advantage of $\bar{F}_{i}^{\rm W}$ becomes particularly critical in aggregative congestion games, especially when the dimension $n$ and population $N$ are large, making such a reduction in computational complexity essential. This advantage is further illustrated through numerical simulations in Subsection~\ref{sec:num_distance}.

Despite its computational advantage in aggregative congestion games, the RGWE remains an approximation of the RGNE. Since the RGWE preserves the stability of the exact solution only when sufficiently close to the RGNE, it is essential to quantify their relationship. The following theorem quantifies a bound, with its proof detailed in Appendix \ref{appen:C}.

\begin{theorem}       \label{thm:distance}
    Under Assumption \ref{asp:convexity_and_nonempty}, an RGWE of the uncertain aggregative congestion game $\mathcal{G}$ is also an $\varepsilon$-RGNE, with 
    \begin{equation}   \nonumber
    \varepsilon = 2 R_{x}^{2} L_{q}  \left(
       \sqrt{\frac{2  L_{q}}{N(\alpha + \beta)}} + \frac{1}{N} \right),
    \end{equation}
    where $R_{x} = \max_{i \in \mathcal{I}} \max_{x_{i} \in \mathcal{X}_{i}} \| x_{i} \|$, while $L_q$, $\alpha$ and $\beta$ denote the Lipschitz, strong convexity, and strong monotonicity constants, respectively, as defined in Assumption~\ref{asp:convexity_and_nonempty}.
\end{theorem}

We observe that the bound $\varepsilon$ in Theorem~\ref{thm:distance} depends on the radius of the local constraint set $\mathcal{X}_i$ and the Lipschitz constant of the per-unit cost $q(\cdot)$, which is straightforward and intuitive. We further note that the dependence on $\alpha$ and $\beta$ follows from the geometry of strong convexity. Recall that an $(\alpha+\beta)$-strongly convex function $f(x)$ satisfies the quadratic growth condition $f(x) \ge f(x^*) + \frac{\alpha+\beta}{2} \| x - x^* \|^2$, which implies $\| x - x^* \| \le \sqrt{\frac{2}{\alpha+\beta} (f(x) - f(x^*)) }$. Given all these parameters fixed, Theorem~\ref{thm:distance} demonstrates that RGWE approximates RGNE, with a bound of order $\mathcal{O}({1 {/} \sqrt{N}})$, where $N$ is the population size. This asymptotic property allows Algorithm \ref{alg} to efficiently approximate the RGNE with reduced computational complexity, a feature particularly advantageous in aggregative congestion games. For instance, with $10^7$ players in charging
control of EVs~\cite{ma2011decentralized}, the distance is of order $3.16 \times 10^{-4}$.

\begin{comment}
\begin{remark}
    A similar result relating vWE and vNE is discussed in \cite{paccagnan2018nash}. However, Theorem \ref{thm:distance} focuses on the relationship between RGWE and RGNE in uncertain games. This is a different setting from \cite[Prop. 2 and Th. 1]{paccagnan2018nash}. Moreover, while \cite[Th. 1.3)]{paccagnan2018nash} assumes that $p_i = 0$ for all $i \in \mathcal{I}$, our result generalizes to the case where $p_i$ is strongly convex.
\end{remark}
\end{comment}

\begin{comment}
    \begin{remark}
    We note the relationship between our result and prior work in \cite[Prop. 2 and Th. 1]{paccagnan2018nash}; however, their results may not be directly applicable to uncertain games. First, as discussed earlier, there is no straightforward solution to $\mathcal{G}$. With the aid of Theorem \ref{thm:robust_trans}, we establish the corresponding variational inequalities. Second, \cite[Th. 1.3)]{paccagnan2018nash} assumes that $p_i = 0$ for all $i \in \mathcal{I}$, while our result generalizes to the case where $p_i$ is strongly convex. In particular, when $p_i = 0$ for all $i \in \mathcal{I}$, \cite[Prop. 2]{paccagnan2018nash} coincides with our Theorem \ref{thm:distance}.        
    \end{remark}
\end{comment}

%
%
%
%
%
%
%
%
%

\section{Application to Robust EV Charging}   \label{sec:simulation}

Based on an EV charging control case study~\cite{ma2011decentralized}, this section presents numerical results to validate the convergence of Algorithm \ref{alg} and the bound in Theorem \ref{thm:distance}, and to demonstrate the computational advantages of $\bar{F}_{i}^{\rm W}$ over $\bar{F}_{i}^{\rm N}$. Additionally, we investigate the impact of uncertainty.

Consider the charging dynamics of each EV $i \in \mathcal{I}$ as a discrete-time linear system $s_i(t+1) = s_i(t) + e_i x_i(t),\ t \in \mathcal{T} =  \{ 0,\dots,T-1\}$, where $s_i(t) \in [0,1]$ denotes the state of charge (SoC) at time interval $t$. The strategy variable $x_i(t) \in [0,1]$ denotes the charging control input, and $e_i > 0$ is the charging efficiency coefficient. At each $t$, the control input $x_i(t)$ should be nonnegative and cannot exceed an upper bound $\overline{x}_i(t)$. Moreover, to reach a desired final SoC, there is a lower bound on the control input over a day, i.e., $\sum_{t \in \mathcal{T}} x_i(t) \ge \underline{u}_i$. This introduces the local constraint for EV $i$'s strategy $x_i = \operatorname{col} (x_i(0),\dots,x_i(T-1))$:
\begin{equation}   \nonumber
    \mathcal{X}_i = \left\{ x_i \in \mathbb{R}^T \middle| 
        0 \leq x_i(t) \leq \bar{x}_i(t), \forall t \in \mathcal{T}, 
        \displaystyle \sum_{t \in \mathcal{T}} x_i(t) \geq \underline{u}_i 
    \right\}.
\end{equation}
The overall power that the grid can afford should not exceed $b$. We model the uncertainties in power grid as uncertainty coefficients $\rho_i$ on the coupling constraints, that is,
$
    \sum_{i \in \mathcal{I}} \rho_i x_i \le b.
$
Finally, the cost function of EV $i$ is modeled as 
\vspace{-0.1cm}
\begin{equation}   \label{eq:simulation}
    J_{i}(x_{i},\sigma(\boldsymbol{x})) =  \underbrace{\frac{1}{2} a_{i} \|  x_{i} - \tilde{x}_i \|^2}_{p_{i}(x_{i}) } + 
    \underbrace{(b_{c} \sigma(\boldsymbol{x}) + c)^\top x_i}_{\quad q(\sigma(\boldsymbol{x}))^\top x_i},
    \vspace{-0.1cm}
\end{equation}
where $\frac{1}{2} a_{i} \|  x_{i} - \tilde{x}_i \|^2$ is the penalty on deviation from the predetermined charging schedule $\tilde{x}_i$, and $b_{c} \sigma(\boldsymbol{x}) + c$ is the energy price vector over the time horizon $T$. Here, $\sigma(\boldsymbol{x}) \in \mathbb{R}^T$ represents the aggregate EV demand, $b_c \in \mathbb{R}$ is the price sensitivity coefficient,  and $c \in \mathbb{R}^T$ represents the base price vector determined by non-EV demand.

\vspace{-0.4cm}
\subsection{Convergence of Algorithm \ref{alg}}  \label{sec:num_convergence}

% \begin{table}[t!]    
% \caption{Numerical parameters}
% \label{tab:num_para}
% \begin{threeparttable}
% \centering
% \setlength{\tabcolsep}{2pt}
% \begin{tabular}{@{} lccccc @{}}
% \toprule
% Parameter & Description & Value \\
% \midrule
% $T$ & Time interval length & 24 \\
% $\{e_i\}_{i \in \mathcal{I}}$ & Charging efficiency & $(0.15,0.15,0.15,0.15,0.15)$ \\
% $\{ s_i(0) \}_{i \in \mathcal{I}} $ & Initial SoC & $(0.10, 0.15, 0.20, 0.25, 0.30)$ \\
% $\{ s_i(T) \}_{i \in \mathcal{I}} $& Desired SoC & $(0.89, 0.75, 0.85, 0.80, 0.90)$ \\
% $\tilde{x}_i(t) $ & Charging schedule & $(s_i(T) - s_i(0))/T,\forall t$ \\
% $\overline{x}_i(t)$ & Upper bound ($x_i(t)$)& 0.2 for all ${ t \in 
% \mathcal{T},  i \in \mathcal{I}}$ \\
% $x_i(0)$ & Initial charging strategy & $\mathcal{U}[0,0.1]$, $\forall i \in \mathcal{I}$ \\
% $\underline{u}_i$ & Lower bound$\left(\sum_{t \in \mathcal{T}} x_i(t)\right)$ & 0.1 for all $i \in \mathcal{I}$ \\
% $D_i$ & Random matrix & Random matrix \\
% $d_i$ & Upper bound($\rho_i$) & $0.5$ for all $i \in \mathcal{I}$ \\
% $b$ & Upper bound $\left(\sum_{i \in \mathcal{I}} \rho_i x_i \right)$ & $5N$ \\
% $a_i$ & Deviation coefficient & $1 + 15 i / N$ \\
% $b_c$ & PEV demand coefficient & 5\\
% $c$ & Non-PEV demand vector & Bidding Zone SE3, Sweden~\cite{NordPool} \\
% \bottomrule
% \end{tabular}
% \end{threeparttable}
% \end{table}   

\begin{table}[t!]    
\caption{Numerical parameters}
\label{tab:num_para}
\begin{threeparttable}
\centering
\setlength{\tabcolsep}{1.5pt}
\begin{tabular}{@{} lccccc @{}}
\toprule
Parameter & Description & Value \\
\midrule
$T$ & Time interval & 24 \\
$\{e_i\}_{i \in \mathcal{I}}$ & Charging efficiency & $(0.85,0.85,0.85,0.85,0.85)$ \\
$\{ s_i(0) \}_{i \in \mathcal{I}} $ & Initial SoC & $(0.10, 0.15, 0.20, 0.25, 0.30)$ \\
$\{ s_i(T) \}_{i \in \mathcal{I}} $& Target SoC & $(0.89, 0.75, 0.85, 0.80, 0.90)$ \\
$\tilde{x}_i(t) $ & Charging schedule & $1.6(s_i(T) - s_i(0))/T,\forall t$ \\
$\overline{x}_i(t)$ & Upper bound on $x_i(t)$ & 0.2 for all ${ t \in 
\mathcal{T},  i \in \mathcal{I}}$ \\
$x_i(0)$ & Initial charging strategy & $\mathcal{U}[0,0.1]$, $\forall i \in \mathcal{I}$ \\
$\underline{u}_i$ & Lower bound on $\sum_{t \in \mathcal{T}} x_i(t)$ & 0.1 for all $i \in \mathcal{I}$ \\
$D_i$ & Uncertainty set matrix & $[I_T,-I_T]^\top$ \\
$d_i$ & Upper bound on $\rho_i$ & $\operatorname{col}( \frac{1}{2} \mathbf{1}_T, - \frac{1}{2}  \mathbf{1}_T),\forall i \in \mathcal{I}$ \\
$b$ & Upper\! bound \!on $\sum_{i \in \mathcal{I}}\! \rho_i x_i$ & $0.5N$ \\
$a_i$ & Deviation coefficient & $2 + 15 i / N$ \\
$b_c$ & EV demand coefficient & 0.1\\
$c$ & Base price vector &Bidding\! Zone \!SE3,\! Sweden~\cite{nordpool}\\
\bottomrule
\end{tabular}
\end{threeparttable}
\end{table} 

\vspace{-0.1cm}
\begin{figure}[t!]
\centering
\includegraphics[width=0.48\textwidth]{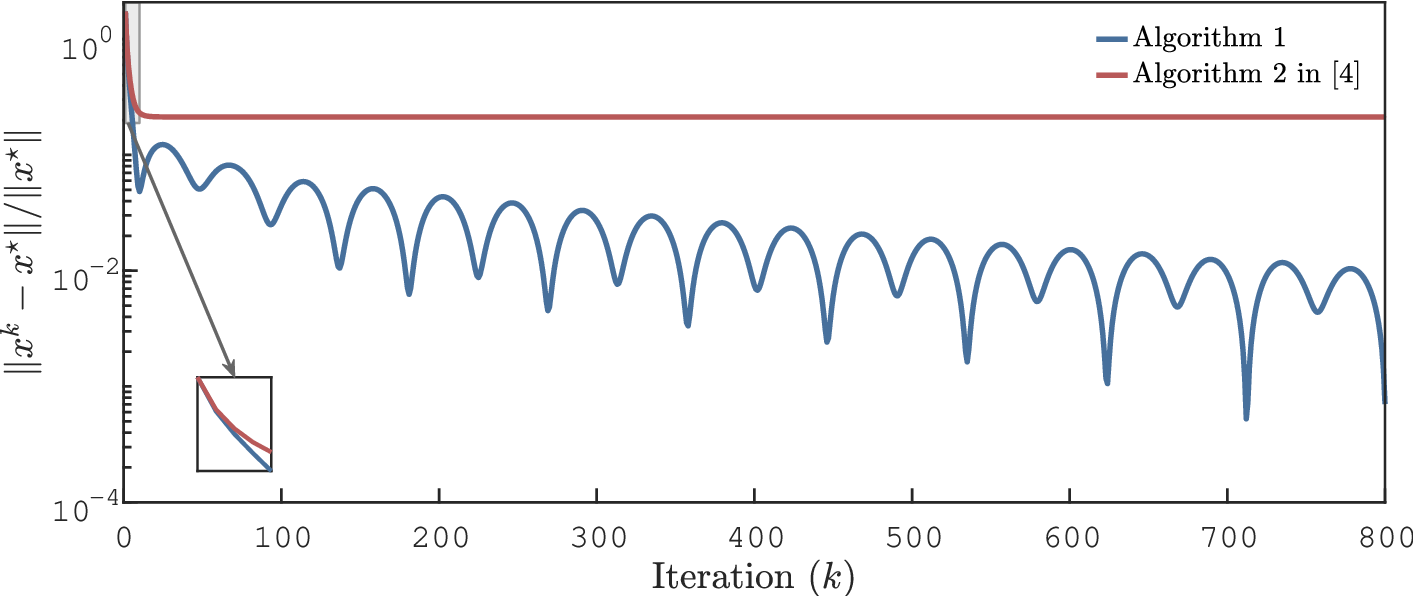}
\caption{Trajectories of the normalized distance $\| \boldsymbol{x}^k - \boldsymbol{x}^\star \| / \| \boldsymbol{x}^\star \|$ as a function of iteration $k$. The proposed Algorithm~\ref{alg} is compared with Algorithm 2 in~\cite{carnevale2024tracking}.}
\label{fig:compare_convergence}

\vspace{-0.5cm}
\end{figure}

In this subsection, we consider $N=5$ EVs specifically to validate the convergence of Algorithm~\ref{alg}. Although model~\eqref{eq:simulation} is usually studied in very large populations (e.g., $N= 10^7$ in~\cite{ma2011decentralized}), we use a small scale here for verification,  deferring the analysis of larger $N$ to Subsection~\ref{sec:num_distance}. We set $\delta = 0.05$ (consistent with~\cite{carnevale2024tracking}) and use parameters from Table~\ref{tab:num_para} to create a deterministic baseline for comparison against~\cite{carnevale2024tracking}. It is worth noting that the GNE seeking algorithm in~\cite{carnevale2024tracking} is modified here to seek the GWE by using the operator $\nabla p_{i}(x_{i}) + q( \sigma(\boldsymbol{x}))$. We consider a communication graph with ring
topology and no self-loops, i.e., $w_{ii} = 0$, for all $i \in \{1,\dots,5\}$.

As shown in Fig.~\ref{fig:compare_convergence}, although Algorithm 2 in~\cite{carnevale2024tracking} converges, it fails to reach the GWE $\boldsymbol{x}^\star$. In contrast, our proposed Algorithm~\ref{alg} successfully converges to $\boldsymbol{x}^\star$. Furthermore, numerical results indicate that the dual variables in \cite{carnevale2024tracking} associated with the coupling constraints fail to maintain non-negativity. This implies that the limit point of Algorithm 2 in~\cite{carnevale2024tracking} is infeasible w.r.t. the coupling constraints, whereas our Algorithm~\ref{alg} guarantees feasibility. Such distinction is theoretically supported by the discussion in Remark~\ref{rmk:alg_compare}. It is also worth noting that, despite lacking a theoretical linear convergence guarantee, the empirical convergence speed of Algorithm~\ref{alg} in the early iterations is highly comparable to Algorithm 2 in~\cite{carnevale2024tracking}. Moreover, our approach is robust to uncertainty in $\rho_i$, as demonstrated in the next subsection.

\vspace{-0.3cm}

\subsection{Impact of Uncertainty}
\vspace{0cm}

\begin{figure}[t!]
\includegraphics[width=0.48\textwidth]{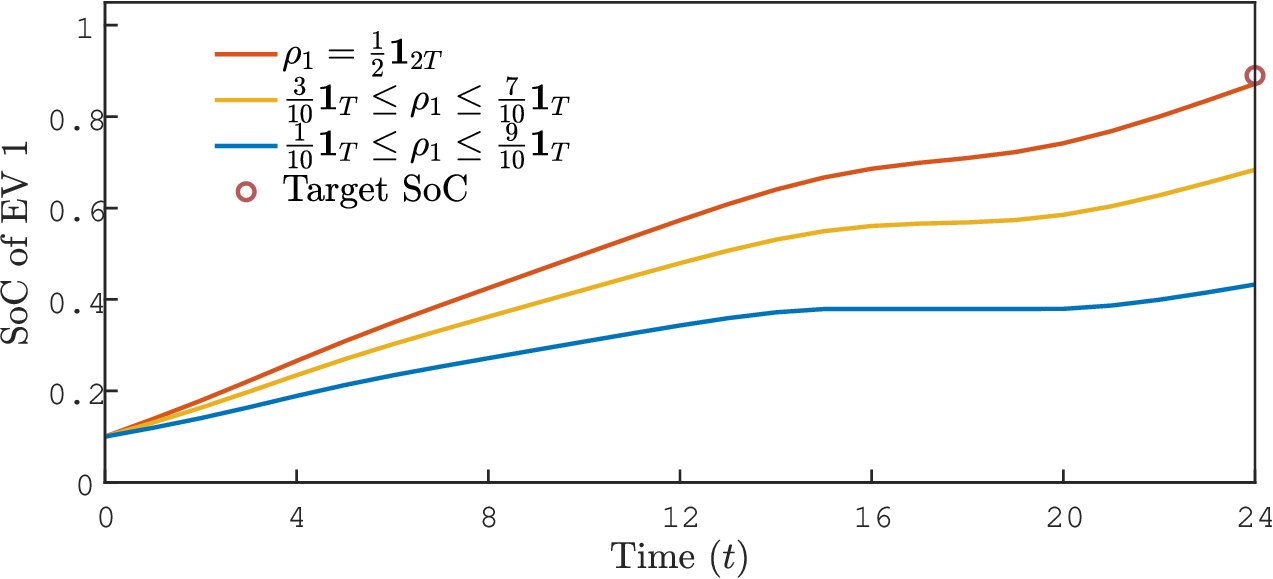}
\caption{SoC of EV 1 for different values of $d_i$ over a day, which determines the size of the box uncertainty set. The red line represents the nominal case where the uncertainty set degenerates to a singleton with $d _i = \frac{1}{2} \operatorname{col}( \mathbf{1}_{T}, - \mathbf{1}_{T} ),\forall i \in \mathcal{I}$, yielding $\rho_1 = \frac{1}{2} \mathbf{1}_{2T}$. The yellow and blue lines correspond to wider uncertainty sets: specifically, $d_i = \operatorname{col}( \frac{7}{10} \mathbf{1}_T, - \frac{3}{10}  \mathbf{1}_T),\forall i \in \mathcal{I}$ results in $ \frac{3}{10} \mathbf{1}_T \le \rho_1 \le \frac{7}{10} \mathbf{1}_T$ (yellow), and $d_i = \operatorname{col}( \frac{9}{10} \mathbf{1}_T, - \frac{1}{10}  \mathbf{1}_T),\forall i \in \mathcal{I}$ results in $\frac{1}{10} \mathbf{1}_T \le \rho_1 \le \frac{9}{10} \mathbf{1}_T$ (blue). Other parameters are detailed in Table~\ref{tab:num_para}.}
\label{fig:SoC_for_EV1}

\vspace{0.3cm}

\includegraphics[width=0.48\textwidth]{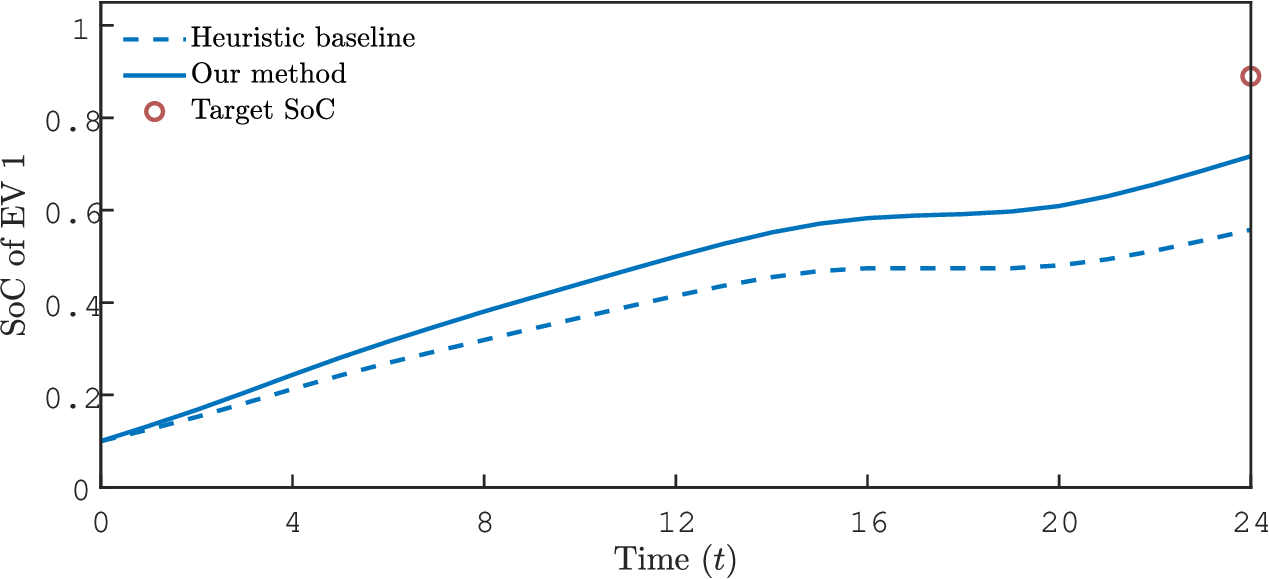}
\caption{SoC trajectories for EV $1$ with two different approaches: the proposed robust method and a baseline strategy. In the baseline strategy, the elements of parameter $\rho_1$ are fixed at $ \frac{4}{5} $ during the day and $\frac{3}{5}$ at night. All simulation parameters other than $d_i$ are listed in Table~\ref{tab:num_para}.}
\label{fig:compare_with_the_trivial_method}
\end{figure}

We examine the SoC trajectories of EV $1$ under different uncertainty levels in Fig.~\ref{fig:SoC_for_EV1}. As expected, the charging rate peaks during off-peak hours (around $t=3$) and drops during peak hours (around $t=17$). Such tendency aligns with established findings in EV charging control literature~\cite{ma2011decentralized,paccagnan2018nash}. Note that setting $D_i = [I_T,-I_T]^\top$ characterizes $P_i$ as a box uncertainty set; specifically, $d _i = \frac{1}{2} \operatorname{col}( \mathbf{1}_{T}, - \mathbf{1}_{T} ),\forall i \in \mathcal{I}$ corresponds to the deterministic case where $\rho_1 = \frac{1}{2} \mathbf{1}_{2T}$.

The charging strategy varies with the uncertainty set size. With a smaller radius, the strategy is aggressive and fully exploits the available grid capacity. In contrast, a larger radius enforces a more conservative approach. This clearly demonstrates the trade-off between robustness and efficiency.

Next, we consider the case of time-varying uncertainty where the uncertainty level is low at night but high during the day. This is modeled by choosing
\[
d_i = \begin{bmatrix}
    \frac{3}{5} \mathbf{1}_8 ; \frac{4}{5} \mathbf{1}_{12} ; \frac{3}{5} \mathbf{1}_4 ; -\frac{2}{5} \mathbf{1}_8  ; -\frac{3}{5} \mathbf{1}_{12} ; -\frac{2}{5} \mathbf{1}_4
\end{bmatrix}, \forall i \in \mathcal{I}.
\]
A heuristic baseline to handle such non-uniform box uncertainty is to fix $\rho_1$ to $\frac{4}{5}$ during the day and $\frac{3}{5}$ at night. As shown in Fig.~\ref{fig:compare_with_the_trivial_method}, our proposed method is less conservative and outperforms this baseline. Moreover, this simple heuristic may fail if $P_i$ becomes a general polyhedron set. These results demonstrate the importance of Theorem~\ref{thm:robust_trans}.

\subsection{RGWE vs. RGNE: Distance and Computational Cost}   \label{sec:num_distance}

\begin{figure}[t!]
\includegraphics[width=0.48\textwidth]{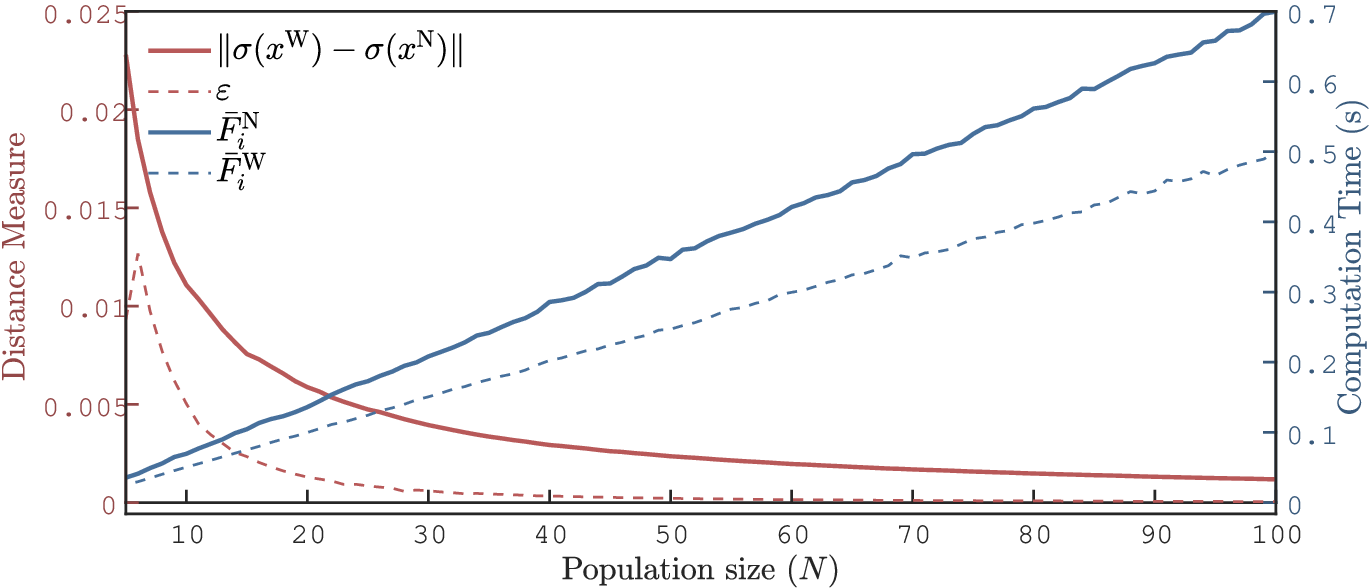}
\caption{Aggregate distance $\| \sigma(\boldsymbol{x}^{\rm W}) - \sigma(\boldsymbol{x}^{\rm N})  \|$ and upper bound $\varepsilon$ (left axis), and computational times for $\bar{F}_{i}^{\rm W}$ and $\bar{F}_{i}^{\rm N}$ (right axis), as a function of  population size $N$.}
\label{fig:computation-complexity}
\end{figure}

We first examine the distance between RGWE and RGNE. Based on \eqref{eq:three_parts}, the upper bound for $\varepsilon$ over the feasible strategy set can be expressed as:
\begin{equation}   \nonumber
\begin{aligned}
    \max_{i \in \mathcal{I}} \Bigg \{
    \ & J_{i}(x^{\rm W}_{i},\sigma(\boldsymbol{x}^{\rm W}))  \\
    & -  \inf_{x_{i} \in \mathcal{C}_{i,\boldsymbol{\rho}}(\boldsymbol{x}_{-i})}
    J_{i}\Bigg(x_{i},\frac{1}{N}  \Bigg(  
    \sum_{j=1,j \neq i}^{N} x_{j}^{\rm W} +  x_{i}  \Bigg) \Bigg)
    \Bigg \}.
\end{aligned}
\end{equation}
To simultaneously evaluate the distance and computational efficiency, we conduct numerical experiments with population sizes varying from $N=1$ to $100$ over time interval $T = 24$. The parameters are generated by periodically repeating the profiles of the first five EVs listed in Table~\ref{tab:num_para}. For each $N$, the operators $\bar{F}_{i}^{\rm W}$ and $\bar{F}_{i}^{\rm N}$ are evaluated $10^4$ times.

The combined results are presented in Fig.~\ref{fig:computation-complexity}. Two key trends are observed: First, the approximation error $\varepsilon$ diminishes as $N \to \infty$, confirming the effectiveness of approximating RGNE via RGWE. Second, regarding computational cost, the time required for the RGNE grows substantially with $N$, whereas the RGWE demonstrates superior scalability with a much slower rate of increase. These empirical findings corroborate the discussions in Subsection~\ref{sec:distance} and Theorem~\ref{thm:distance}.

\vspace{-0.3cm}
\section{Conclusion}  \label{sec:conclusion}
This paper analyzed the RGWE in aggregative congestion games subject to uncertain coupling constraints. We demonstrated that the RGWE offered a tractable and robust formulation for such environments, and verified its close proximity to the RGNE in large populations. By leveraging singular perturbation analysis and dynamic tracking techniques, we proposed a fully distributed algorithm for RGWE seeking. We identified the investigation of uncertain nonlinear constraints as a natural extension, broadening the applicability of our framework beyond the affine case. Finally, we suggested that future research explore alternative game-theoretic models and equilibrium concepts to achieve a superior trade-off between computational efficiency and equilibrium accuracy.

\appendix
\section{Appendix}
\subsection{Proof of Theorem \ref{thm:robust_trans}}   \label{appen:A}    
    Sufficiency. Suppose 
    $(\boldsymbol{x},\boldsymbol{\gamma})$ 
    is feasible with respect to the condition $(\ref{eq:augmented_game})$. 
    We have
    \begin{equation}   \label{eq:constraint}
       \sum_{j=1}^{N}d_{j}^\top \gamma_{j} \le b, \ D_{i}^\top \gamma_{i} - x_{i}=\mathbf{0}_{n},\ \forall i \in \mathcal{I}.
    \end{equation}
    Then, it is natural that the following
    \begin{equation}      \label{eq:min_constraint}
        \begin{aligned}
            \sum_{j=1}^{N}\min_{\gamma_{j}} \ 
            d_{j}^\top \gamma_{j} \le b, 
            \ D_{i}^\top\gamma_{i} - x_{i} = \mathbf{0}_{n}, \forall i \in \mathcal{I}
        \end{aligned} 
    \end{equation}
    holds. 
    We treat each subproblem in (\ref{eq:min_constraint}) as an independent optimization problem:
    \begin{equation}    \label{eq:independent_min_opt}
        \begin{aligned}
            & \min_{\gamma_{i} \in \mathbb{R}^{m}_{+}} \ d_{i}^\top  \gamma_{i}
            & \begin{array}{r@{\quad}r@{}l@{\quad}l}
            \text{s.t.} \  D_{i}^\top \gamma_{i}  - x_{i} = \mathbf{0}_{n}.
            \end{array}
        \end{aligned}
    \end{equation}
    Note that $\gamma_{i}$ is the decision variable of problem~\eqref{eq:independent_min_opt} and $d_{i}$ is a constant vector. 
    Thus, the Lagrange dual of problem~\eqref{eq:independent_min_opt} is 
    $
            \max_{\rho _{i}} x_{i}^\top \rho_{i},\
            \text{s.t.}\ D_{i} \rho_{i} \le d_{i},
    $
    and condition (\ref{eq:constraint}) becomes
    $
        \sum_{i=1}^{N}\max_{\rho_{i} \in P_{i}} \rho_{i}^\top x_{i}
        \le b, \ \ 
        \text{s.t.} \ D_{i} \rho_{i} \le d_{i},
    $
    whose maximum could be removed based on \cite[Sec. 2.2]{bertsimas2011theory}, i.e.,
    \begin{equation}   \nonumber
        \sum_{i=1}^{N}\rho_{i}^\top x_{i}
        \le b, \ \ 
        \text{s.t.} \ D_{i} \rho_{i} \le d_{i},\ \rho _{i}\in P_{i},\ x_{i} \in \mathcal{X}_{i}.
    \end{equation}
    Therefore,  any feasible solution satisfying the condition (\ref{eq:augmented_game}) also satisfies the coupling constraints \eqref{eq:coupling_constraints_overall}.

    Necessity. According to the coupling constraints \eqref{eq:coupling_constraints_overall}, the maximum of the left side of the inequality satisfies
    \begin{equation}    \label{eq:max_constraint}
        \begin{aligned}
            \max_{\boldsymbol{\rho} \in \boldsymbol{P}} \ 
            \sum_{i=1}^{N}\rho_{i}^\top x_{i}
            =\sum_{i=1}^{N} \max_{\rho _{i} \in P_{i}}\rho_{i}^\top x_{i}
            \le b.
        \end{aligned} 
    \end{equation}
    Since each subproblem in (\ref{eq:max_constraint}) can be considered independently, we can reformulate the original expression as a collection of isolated optimization problems. 
    Moreover, it follows that
    \begin{equation}    \label{eq:sub_max_opt}
        \begin{aligned}
        & \max_{\rho _{i}}  \ x_{i}^\top  \rho_{i}
        & \begin{array}{r@{\quad}r@{}l@{\quad}l}
        \text{s.t.}\ D_{i} \rho_{i} \le d_{i},
        \end{array}
    \end{aligned}   
    \end{equation}
    where $\rho_{i}$ is the decision variable of subproblem~\eqref{eq:sub_max_opt}, with $x_{i}$ being a constant vector. 
    Now, introduce a Lagrangian function 
    with an auxiliary multiplier $\gamma_{i}\in \mathbb{R}^{m}_{+}$, i.e.,
    $
            \mathcal{L}(\rho_{i},\gamma_{i})
            = -x_{i}^\top \rho_{i}+\gamma_{i}^\top (D_{i} \rho_{i} - d_{i}).
    $
    Under Assumption \ref{asp:convexity_and_nonempty}, the polyhedron $P_{i}$
    has a nonempty interior, and hence Slater’s condition is satisfied in optimization problem (\ref{eq:sub_max_opt}).
    Consequently, the dual gap vanishes, and the Lagrangian function $\mathcal{L}(\rho_{i},\gamma_{i})$ can be expressed as
    \begin{equation}  \nonumber
        \begin{aligned}
        & \min_{\gamma_{i}} \ d_{i}^\top  \gamma_{i}
        & \begin{array}{r@{\quad}r@{}l@{\quad}l}
        \text{s.t.}\ D_{i}^\top \gamma_{i} - x_{i}=\mathbf{0}_{n}, \ \gamma_{i} \ge \mathbf{0}_{m}.
        \end{array}
        \end{aligned}  
    \end{equation}
    Therefore, the inequality (\ref{eq:max_constraint}) under the worst case can be rewritten as
    \begin{equation} \label{eq:min_constraint_nec}
        \begin{aligned}
            \sum_{i=1}^{N}\min_{\gamma_{i}\in \Gamma_{i}} \ 
            d_{i}^\top\gamma_{i}  
            \le b,
        \end{aligned} 
    \end{equation}
    where
    $\Gamma_{i}=\{ \gamma_{i}\in \mathbb{R}^{m}_{+} \mid D_{i}^\top \gamma_{i} - x_{i}=\mathbf{0}_{n} \}$. 
    Following \cite[Sec. 2.2]{bertsimas2011theory}, the minimization on the left-hand side of (\ref{eq:min_constraint_nec}) can be removed provided that there exists at least one $\boldsymbol{\gamma}$ satisfying the constraints in (\ref{eq:min_constraint_nec}), namely, 
    $
            \sum_{i=1}^{N}   d_{i}^\top \gamma_{i} 
            \le b , \ D_{i}^\top \gamma_{i} - x_{i}=\mathbf{0}_{n},\ 
            \forall i\in \mathcal{I},$
    where $x_{i}\in \mathcal{X}_{i}$ and $\gamma_{i}\in \mathbb{R}^{m}_{+}$ are decision variables. 
    This is exactly the expression of (\ref{eq:augmented_game}). 
    Hence, feasibility w.r.t. \eqref{eq:augmented_game} implies feasibility w.r.t. \eqref{eq:coupling_constraints_overall}, and the proof follows.   

\subsection{Proof of Theorem \ref{thm:convergence}}    \label{appen:B}

The proof consists of six steps. We begin by applying a coordinate transformation to the fast dynamics~\eqref{eq:alg:e}. Using this transformation and the doubly stochastic property of $W$ according to Assumption~\ref{asp:convexity_and_nonempty}, we rewrite the algorithm \eqref{eq:algorithm_compact} into the form of~\eqref{eq:trans_sin_per_sys} to analyze under singular perturbation theory. Next, we show that the transformed fast dynamics~\eqref{eq:trans_sin_per_sys:b} is asymptotically stable at the origin. In Step~4, we establish that the equilibrium of the transformed slow dynamics~\eqref{eq:trans_sin_per_sys:a} corresponds to an RGWE of $\mathcal{G}$. Step 5 proves the convergence of \eqref{eq:trans_sin_per_sys:a}, and finally, by LaSalle’s invariance principle, we conclude that the entire system~\eqref{eq:trans_sin_per_sys} converges to an RGWE.

\textbf{Step 1.} Denote the error variable $\tilde{\boldsymbol{\sigma}}^k := \hat{\boldsymbol{\sigma}}^k - \boldsymbol{y}^k$. We then have $\tilde{\boldsymbol{\sigma}}^{k+1} = \hat{\boldsymbol{\sigma}}^{k+1} - \boldsymbol{y}^{k+1} = (W \otimes I_{n+m})\hat{\boldsymbol{\sigma}}^k - \boldsymbol{y}^k$. Expanding this term yields $(W \otimes I_{n+m})(\tilde{\boldsymbol{\sigma}}^k + \boldsymbol{y}^k) - \boldsymbol{y}^k$, which simplifies to $(W \otimes I_{n+m})\tilde{\boldsymbol{\sigma}}^k - (L \otimes I_{n+m})\boldsymbol{y}^k$.

    Given the initialization $\hat{\boldsymbol{\sigma}}^{0} = \boldsymbol{y}^{0}$, 
    $\tilde{\boldsymbol{\sigma}}^{0} $ 
    satisfies $\frac{1}{N}\mathbf{1}_{N}^{\top} \otimes I_{n+m} \tilde{\boldsymbol{\sigma}}^{0} = \mathbf{0}_{n+m} $. 
    Introduce coordinates as follows  
    \begin{equation}     \label{eq:coordinates_change}
    \begin{aligned}
        \begin{bmatrix}
            \tilde{\boldsymbol{\sigma}}_{-} \\
            \tilde{\boldsymbol{\sigma}}_{\perp}
        \end{bmatrix}
        : & = \begin{bmatrix}
            \frac{1}{N}\mathbf{1}_{N}^{\top} \otimes I_{n+m} \\
            R_{n+m}^{\top} 
        \end{bmatrix}   \tilde{\boldsymbol{\sigma}},
    \end{aligned}
    \end{equation}
    which implies $
    \tilde{\boldsymbol{\sigma}} = \mathbf{1}_{N} \otimes I_{n+m} \tilde{\boldsymbol{\sigma}}_{-} 
    + R_{n+m} \tilde{\boldsymbol{\sigma}}_{\perp}$, with $R_{n+m} \in \mathbb{R}^{N(n+m) \times (N - 1)(n+m)} $, 
    $R_{n+m}^{\top } R_{n+m} = I_{(N-1)(n + m)} $, 
    $R_{n+m} R_{n+m}^{\top } = I_{N(n + m)}  - \frac{1}{N} \mathbf{1}_{N} \mathbf{1}_{N} ^{\top} \otimes  I_{n+m} $,  
    and $R_{n+m}^{\top } \mathbf{1}_{N} \otimes I_{n + m} = \mathbf{0}_{(N - 1)(n+m) \times (n+m)}  $.
    Here, $\tilde{\boldsymbol{\sigma}}_{-} \in \mathbb{R}^{n+m}$ and 
    $\tilde{\boldsymbol{\sigma}}_{\perp} \in \mathbb{R}^{(N-1)(n+m)}$.
    The mean value $\tilde{\boldsymbol{\sigma}}_{-}$ satisfies
    \begin{align*}
    & \tilde{\boldsymbol{\sigma}}-^{k+1} \\ 
    & = \frac{1}{N}(\mathbf{1}N^\top \otimes I_{n+m}) \big( (W \otimes I_{n+m})\tilde{\boldsymbol{\sigma}}^k - (L \otimes I_{n+m})\boldsymbol{y}^k \big) \nonumber \\
    &= \frac{1}{N}(\mathbf{1}N^\top \otimes I_{n+m})\tilde{\boldsymbol{\sigma}}^k = \tilde{\boldsymbol{\sigma}}_-^k,
    \end{align*}
    where the second equality holds since $\mathbf{1}_N^\top W = \mathbf{1}_N^\top$ and $\mathbf{1}_N^\top L = \mathbf{0}_N$.
    Note that the initialization implies  
    $\tilde{\boldsymbol{\sigma}}_{-} ^{k + 1} = \tilde{\boldsymbol{\sigma}}_{-} ^{k} = \mathbf{0}_{n+m} $, 
    for all $k \ge 0$. Omitting the null dynamics $\tilde{\boldsymbol{\sigma}}_{-} ^{k}$ (i.e., $\tilde{\boldsymbol{\sigma}}^k = R_{n+m}\tilde{\boldsymbol{\sigma}}_\perp^k$) and applying \eqref{eq:coordinates_change}, the remaining dynamics reduce to
    \begin{equation}         \nonumber
    \begin{aligned}
        \tilde{\boldsymbol{\sigma}}_{\perp}^{k+1}  = & R_{n+m}^{\top} \tilde{\boldsymbol{\sigma}}^{k+1} \\
        = & R_{n+m}^{\top} (W \otimes I_{n+m}) R_{n+m} \tilde{\boldsymbol{\sigma}}_{\perp}^{k} 
        - R_{n+m}^{\top} ( L \otimes I_{n+m} )\boldsymbol{y}^{k}.
    \end{aligned}
\end{equation}

\textbf{Step 2.} Denote $\chi^{k}  := \operatorname{col}(
    \boldsymbol{y}^{k}  ,  \boldsymbol{\nu}^{k} , \boldsymbol{\mu}^{k} ,  \boldsymbol{\lambda}^{k} )
    $, 
$\omega^{k} := 
    \tilde{\boldsymbol{\sigma}}_{\perp}^{k}  $
, then we can rewrite \eqref{eq:algorithm_compact} as 
\begin{subequations}     \label{eq:singu_per_sys}
\begin{align}
    \chi^{k + 1} &=  f(\chi^{k}, \omega^{k}, \delta), \label{eq:singu_per_sys:a}   \\
    \omega^{k + 1} & = g(\chi^{k}, \omega^{k}), \label{eq:singu_per_sys:b}
\end{align}
\end{subequations}
in which $f(\chi^{k}, \omega^{k}, \delta) = $
$$
\begin{bmatrix}
    \Pi_{\boldsymbol{\mathcal{Y}}} \left[ \boldsymbol{y}^{k}
        - \delta \left( \boldsymbol{\phi}(\boldsymbol{y}^{k}, \boldsymbol{y}^{k} +  R_{n+m} \tilde{\boldsymbol{\sigma}}_{\perp}^{k} ) + \boldsymbol{A} \boldsymbol{\mu}^{k} + \boldsymbol{B}\boldsymbol{\lambda}^{k} \right) \right] \\
        \boldsymbol{\nu}^{k}  + \delta L \boldsymbol{\mu}^{k} \\
        \Pi_{\mathbb{R}_{+}^{N}}  
        \left[ \boldsymbol{\mu}^{k} 
        + \delta (\boldsymbol{A}^{\top} \boldsymbol{y}^{k} - \boldsymbol{b}  ) 
        - \delta L \boldsymbol{\mu}^{k} - \delta L \boldsymbol{\nu}^{k} \right] \\
        \boldsymbol{\lambda}^{k}
        + \delta \boldsymbol{B}^{\top} \boldsymbol{y}^{k}
\end{bmatrix}$$
and
\begin{equation}  \nonumber
\begin{aligned}
    g(\chi^{k}, \omega^{k}) & = 
    R_{n+m}^{\top} (W \otimes I_{n+m}) R_{n+m} \omega^{k} \\
    - R_{n+m}^{\top} ( L & \otimes I_{n+m} )
    ( [ I_{N(n + m)} \ \mathbf{0}_{N(n+m) \times N(n+2)} ] \chi^{k}).
\end{aligned}
\end{equation}

Here, we consider (\ref{eq:singu_per_sys:b}) as the fast system and (\ref{eq:singu_per_sys:a}) as the slow system in the framework of singular perturbation. When $\delta$ is sufficiently small, (\ref{eq:singu_per_sys:a}) will remain unchanged while (\ref{eq:singu_per_sys:b}) will evolve at a relatively quick speed. Thus, singular perturbation is an ideal tool to deal with such dynamics. First, let us investigate the fast dynamics (\ref{eq:singu_per_sys:b}). When the slow state $\chi ^{k} = \chi$ is fixed, let $\tilde{\boldsymbol{\sigma}}_{\perp} = - R_{n+m}^{\top} \boldsymbol{y}$. Utilizing the standard properties $R_{n+m}R_{n+m}^\top = I_{N(n+m)} - \frac{1}{N}\mathbf{1}_N\mathbf{1}_N^\top \otimes I_{n+m}$, $W\mathbf{1}_N = \mathbf{1}_N$, and $W + L = I_N$, the RHS of (\ref{eq:singu_per_sys:b}) is
\begin{align*}  
    & R_{n+m}^{\top} (W \otimes I_{n+m}) R_{n+m} \tilde{\boldsymbol{\sigma}}_{\perp} - R_{n+m}^{\top} ( L \otimes I_{n+m} )\boldsymbol{y} \\
    % %
    % %
    % = & - R_{n+m}^{\top}  (W \otimes I_{n+m}) R_{n+m} R_{n+m}^{\top} \boldsymbol{y}
    % %
    % - R_{n+m}^{\top} ( L \otimes I_{n+m} )\boldsymbol{y} \\
    % %
    % %
    % = & - R_{n+m}^{\top} 
    % %
    % \left[ (W \otimes I_{n+m}) R_{n+m} R_{n+m}^{\top} + 
    % %
    % ( L \otimes I_{n+m} ) \right] 
    % %
    % \boldsymbol{y}  \\
    % %
    % %
    % = & - R_{n+m}^{\top} 
    % %
    % \Big[ (W \otimes I_{n+m})  - \frac{1}{N} W \mathbf{1}_{N} \mathbf{1}_{N} ^{\top} \otimes  I_{n+m} \\
    % %
    % %
    % & + ( L \otimes I_{n+m} ) \Big] \boldsymbol{y} \\
    % %
    % %
    % = &  - R_{n+m}^{\top} 
    % %
    % \left[ (W - \frac{1}{N} \mathbf{1}_{N} \mathbf{1}_{N}^{\top} + L )\otimes I_{n+m}  \right] \boldsymbol{y} \\
    % %
    % %
    % = & - R_{n+m}^{\top} 
    % %
    % \left[ (I_{N} - \frac{1}{N} \mathbf{1}_{N} \mathbf{1}_{N}^{\top} )\otimes I_{n+m}  \right] \boldsymbol{y} \\
    %
    %
    = & - R_{n+m}^{\top} \boldsymbol{y} = \tilde{\boldsymbol{\sigma}}_{\perp}.
\end{align*}
Thus, the equilibrium of (\ref{eq:singu_per_sys:b}) is 
$\omega^{k + 1} = \omega^{k} = 
    h(\chi) = 
    - R_{n+m}^{\top} \boldsymbol{y} 
$, satisfying
$
    h(\chi) = g( \chi, h(\chi) ).
$
Consider the error between $\omega^{k} $ and its equilibrium $h(\chi)$, defined as $\psi^{k} := \omega^{k} - h(\chi)$ for a fixed $\chi$. By substituting $\omega^{k} = \psi^{k} + h(\chi)  $ into (\ref{eq:singu_per_sys:b}), we obtain
$
        \psi^{k + 1 } - R_{n+m}^{\top} \boldsymbol{y} = R_{n+m}^{\top} (W \otimes I_{n+m}) R_{n+m} 
        (\psi^{k} - R_{n+m}^{\top} \boldsymbol{y} )  
        - R_{n+m}^{\top}  ( L \otimes I_{n+m} )
        \boldsymbol{y}.
$ 
Noting that $I_{N(n+m)} - R_{n+m} R_{n+m}^\top = \frac{1}{N}\mathbf{1}_N\mathbf{1}_N^\top \otimes I_{n+m}$ and $R_{n+m}^\top (\mathbf{1}_N \otimes I_{n+m}) = \mathbf{0}_{(N - 1)(n+m) \times (n+m)}$, the terms involving $\boldsymbol{y}$ cancel out, yielding
$
        \psi^{k + 1}
        =  R_{n+m}^{\top} (W \otimes I_{n+m}) R_{n+m} \psi^{k}.
$

Now, \eqref{eq:singu_per_sys} could be rewritten as 
\begin{subequations}  \label{eq:trans_sin_per_sys}
    \begin{align}
    \chi^{k + 1} &=  f(\chi^{k}, \psi^{k} + h(\chi^{k}), \delta),  \label{eq:trans_sin_per_sys:a}\\
    \psi^{k + 1} & = R_{n+m}^{\top} (W \otimes I_{n+m}) R_{n+m} \psi^{k}.\label{eq:trans_sin_per_sys:b}
\end{align}
\end{subequations}

\textbf{Step 3.} We consider (\ref{eq:trans_sin_per_sys:b}) as a linear system. Notice that $W$ is doubly stochastic. Together with the coordinates defined in (\ref{eq:coordinates_change}), the matrix $S : = R_{n+m}^{\top} (W \otimes I_{n+m}) R_{n+m}$ is a Schur matrix, i.e., all eigenvalues lie strictly inside the unit circle. Thus, the fast (error) dynamics (\ref{eq:trans_sin_per_sys:b}) are asymptotically stable at the origin, meaning $\psi^{\star} = 0$ for a sufficiently small $\delta$. Equivalently, it means that fast dynamics (\ref{eq:singu_per_sys:b}) converges to its equilibrium $\omega^{k} = h(\chi^{k})$. In addition, there exists matrix $P_{1}=P_{1}^{\top} \succ 0$ satisfying the Lyapunov equation
$
    S^{\top} P_{1} S  - P_{1} = -I.
$

Choose the Lyapunov function of the fast dynamics as 
$
    V_{f}(\psi) = \psi^{\top} P_{1} \psi ,
$
and denote the difference of $V_{f}$ along the trajectories along (\ref{eq:trans_sin_per_sys}) as $\Delta V_{f} (\psi^{k})  :=  V_{f}(\psi^{k+1}) -  V_{f}(\psi^{k}) $,
\begin{equation}   \label{eq:lyap_for_fast_dyna}
    \Delta V_{f} (\psi^{k}) 
    = (\psi ^{k})^{\top} (S^{\top} P_{1} S - P_{1} ) \psi ^{k}
    = - \|  \psi ^{k}   \|^{2}.
\end{equation}

\textbf{Step 4.} Now, it is time to study the slow dynamics (\ref{eq:trans_sin_per_sys:a}). Set $\omega^{k} = h(\chi^{k})$, then (\ref{eq:trans_sin_per_sys:a}) becomes
\begin{subequations}    \label{eq:slow_dyna}
    \begin{align}
        \boldsymbol{y}^{k+1}  = & 
        \Pi_{\boldsymbol{\mathcal{Y}}} \Big[ \boldsymbol{y}^{k}
        - \delta \Big( \boldsymbol{\phi}(\boldsymbol{y}^{k},  \frac{1}{N} \mathbf{1}_{N} \mathbf{1}_{N} ^{\top} \otimes  I_{n+m}\boldsymbol{y}^{k} )  \notag  \\
        & \quad \quad + \boldsymbol{A} \boldsymbol{\mu}^{k} + 
        \boldsymbol{B}\boldsymbol{\lambda}^{k} \Big) \Big] ,  \label{eq:slow_dyna:a}   \\ 
         \boldsymbol{\nu}^{k + 1} = & \boldsymbol{\nu}^{k}  + \delta L \boldsymbol{\mu}^{k},  \label{eq:slow_dyna:b}  \\
        \boldsymbol{\mu}^{k + 1} = & \Pi_{\mathbb{R}_{+}^{N}}  
        \left[ \boldsymbol{\mu}^{k} 
        + \delta (\boldsymbol{A}^{\top} \boldsymbol{y}^{k} - \boldsymbol{b}  ) 
        - \delta L \boldsymbol{\mu}^{k} - \delta L \boldsymbol{\nu}^{k} \right],  \label{eq:slow_dyna:c}    \\
         \boldsymbol{\lambda}^{k + 1} =  & \boldsymbol{\lambda}^{k} 
        + \delta  \boldsymbol{B}^{\top} \boldsymbol{y}^{k}. \label{eq:slow_dyna:d}  
    \end{align}
\end{subequations}
Note that $ \frac{1}{N} \mathbf{1}_{N} \mathbf{1}_{N} ^{\top} \otimes  I_{n+m} \boldsymbol{y}^{k}  = \mathbf{1}_{N} \otimes \frac{1}{N} \sum_{i=1}^{N} y_{i}^{k} =  \mathbf{1}_{N} \otimes \bar{\sigma}(\boldsymbol{y}^{k}) $. 
Thus, $\boldsymbol{\phi}(\boldsymbol{y}^{k},  \frac{1}{N} \mathbf{1}_{N} \mathbf{1}_{N} ^{\top} \otimes  I_{n+m}\boldsymbol{y}^{k} )
= \boldsymbol{\phi}(\boldsymbol{y}^{k}, \mathbf{1}_{N} \otimes \bar{\sigma}(\boldsymbol{y}^{k}) )
= \bar{\boldsymbol{F}}^{\rm W}(\boldsymbol{y}^{k})$. 
Hence, the equilibrium of (\ref{eq:slow_dyna:a}) satisfies
\begin{equation}    \nonumber
    \boldsymbol{y}^{\star}  = 
        \Pi_{\boldsymbol{\mathcal{Y}}} \left[ \boldsymbol{y}^{\star}
        - \delta \left( \bar{\boldsymbol{F}}^{\rm W}(\boldsymbol{y}^{\star}) + \boldsymbol{A} \boldsymbol{\mu}^{\star} + \boldsymbol{B}\boldsymbol{\lambda}^{\star} \right) \right].
\end{equation}

The equilibrium of (\ref{eq:slow_dyna:b}) is given by $L \boldsymbol{\mu}^{*} = \mathbf{0}_{N}$, which is exactly (\ref{eq:KKT:d}). Now, let us check the complementary slackness. At equilibrium, (\ref{eq:slow_dyna:c}) must satisfy
\begin{equation}  \nonumber
    \boldsymbol{\mu}^{\star} = \Pi_{\mathbb{R}_{+}^{N}}  
        \left[ \boldsymbol{\mu}^{\star} 
        + \delta (\boldsymbol{A}^{\top} \boldsymbol{y}^{\star} - \boldsymbol{b}  ) 
        - \delta L \boldsymbol{\mu}^{\star} - \delta L \boldsymbol{\nu}^{\star} \right].
\end{equation}
By Proposition \ref{pps:property_of_proj} and $L \boldsymbol{\mu}^{*} = \mathbf{0}_{N}$, it is equivalent to, for all $\boldsymbol{\mu} \in \mathbb{R}_{+}^{N}$
\begin{equation}    \label{eq:prove_equi_to_KKT}
    - (\boldsymbol{A}^{\top} \boldsymbol{y}^{\star} - \boldsymbol{b} )^{\top}  \boldsymbol{\mu}^{\star}
    + ( \boldsymbol{A}^{\top} \boldsymbol{y}^{\star} - \boldsymbol{b}  - L \boldsymbol{\nu}^{\star} )^{\top}
    \boldsymbol{\mu} \le 0.
\end{equation}

First, assume that $\boldsymbol{A}^{\top} \boldsymbol{y}^{\star} - \boldsymbol{b}  - L \boldsymbol{\nu}^{\star} > \mathbf{0}_{N}$, then there exists a sufficiently large $\boldsymbol{\mu}$, such that (\ref{eq:prove_equi_to_KKT}) is not satisfied. This contradiction implies $\boldsymbol{A}^{\top} \boldsymbol{y}^{\star} - \boldsymbol{b}  - L \boldsymbol{\nu}^{\star} \le \mathbf{0}_{N}$. Left-multiplying this inequality with $\mathbf{1}_{N}^{\top}$, together with $\mathbf{1}_{N}^{\top} L = \mathbf{0}_{N}$, we have 
$
    \mathbf{1}_{N}^{\top}  (\boldsymbol{A}^{\top} \boldsymbol{y}^{\star} - \boldsymbol{b}) \le 0.
$
Moreover, let $\boldsymbol{\mu} = \boldsymbol{0}_{N}$ and $2\boldsymbol{\mu}^{\star}$, with $\mathbf{1}_{N}^{\top} L = \mathbf{0}_{N}$, (\ref{eq:prove_equi_to_KKT}) yields
$
    (\boldsymbol{A}^{\top} \boldsymbol{y}^{\star} - \boldsymbol{b} )^{\top}  \boldsymbol{\mu}^{\star}
    = 0.
$

With these analyses above, the equilibrium of (\ref{eq:slow_dyna}), $(\boldsymbol{y}^{\star}, \boldsymbol{\nu}^{\star}, \boldsymbol{\mu}^{\star}, \boldsymbol{\lambda}^{\star})$, satisfies 
\begin{subequations}       \label{eq:KKT_pre}
    \begin{align}
        \boldsymbol{y}^{\star} & = 
        \Pi_{\boldsymbol{\mathcal{Y}}} \left[ \boldsymbol{y}^{\star}
        - \delta \left( \bar{\boldsymbol{F}}^{\rm W}(\boldsymbol{y}^{\star}) + \boldsymbol{A} \boldsymbol{\mu}^{\star} + \boldsymbol{B}\boldsymbol{\lambda}^{\star} \right) \right], \label{eq:KKT_pre:a} \\
       0 & \ge \mathbf{1}_{N}^{\top} (\boldsymbol{A}^{\top} \boldsymbol{y}^{\star} - \boldsymbol{b}),  \quad
        0 = ( \boldsymbol{A}^{\top} \boldsymbol{y}^{\star} - \boldsymbol{b} )^{\top}  \boldsymbol{\mu}^{\star},   \\
       \mathbf{0}_{Nn} & =  \boldsymbol{B}^{\top} \boldsymbol{y}^{\star}, \\
       \mathbf{0}_{N} & =  L \boldsymbol{\mu}^{\star}.
    \end{align}
\end{subequations}
By Proposition \ref{pps:property_of_proj}, for the nonempty closed convex set $\boldsymbol{\mathcal{Y}}$, (\ref{eq:KKT_pre:a}) is equivalent to the stationarity condition (\ref{eq:KKT:a}), thus (\ref{eq:KKT_pre}) is equivalent to the KKT condition (\ref{eq:KKT}).
Therefore, $(\boldsymbol{y}^{\star}, \boldsymbol{\nu}^{\star}, \boldsymbol{\mu}^{\star}, \boldsymbol{\lambda}^{\star})$
is an RGWE of $\mathcal{G}$.

\textbf{Step 5.} Now, it is time to investigate the slow dynamics (\ref{eq:slow_dyna}). 
Define a quadratic Lyapunov function candidate as
\begin{equation}    \nonumber
    V_{s}( \chi ) = \| \boldsymbol{y} - \boldsymbol{y}^{\star} \|^{2} 
    + \| \boldsymbol{\nu}^{k} - \boldsymbol{\nu}^{\star} \|^{2}
    + \| \boldsymbol{\mu}^{k} - \boldsymbol{\mu}^{\star} \|^{2} 
    + \| \boldsymbol{\lambda} - \boldsymbol{\lambda}^{\star} \|^{2},
\end{equation}
where $(\boldsymbol{y}^{\star}, \boldsymbol{\nu}^{\star}, \boldsymbol{\mu}^{\star}, \boldsymbol{\lambda}^{\star})$ is the equilibrium of reduced system \eqref{eq:slow_dyna},
then we have
\begin{equation}     \label{eq:lyap_for_slow_dyna}
    \begin{aligned}
        V_{s}( \chi ^{k + 1}) =& \| \boldsymbol{y}^{k + 1} - \boldsymbol{y}^{\star} \|^{2} 
    + \| \boldsymbol{\nu}^{k} - \boldsymbol{\nu}^{\star} \|^{2}   \\
    &+ \| \boldsymbol{\mu}^{k + 1} - \boldsymbol{\mu}^{\star} \|^{2}   
    + \| \boldsymbol{\lambda}^{k + 1} - \boldsymbol{\lambda}^{\star} \|^{2}.
    \end{aligned}
\end{equation}

Let us check the four parts of \eqref{eq:lyap_for_slow_dyna}. 
The squared distance
\begin{align*}  
        & \| \boldsymbol{y}^{k + 1} - \boldsymbol{y}^{\star} \|^{2} \\
        & =  \Big\lVert  \Pi_{\boldsymbol{\mathcal{Y}}} \left[ \boldsymbol{y}^{k}
        - \delta \left( \bar{\boldsymbol{F}}^{\rm W}(\boldsymbol{y}^{k} ) + 
        \boldsymbol{A} \boldsymbol{\mu}^{k} + 
        \boldsymbol{B}\boldsymbol{\lambda}^{k} \right) \right]  \\
        &\quad - \Pi_{\boldsymbol{\mathcal{Y}}} \left[ \boldsymbol{y}^{\star}
        - \delta \left( \bar{\boldsymbol{F}}^{\rm W}(\boldsymbol{y}^{\star}) + \boldsymbol{A} \boldsymbol{\mu}^{\star} + \boldsymbol{B}\boldsymbol{\lambda}^{\star} \right) \right]  
        \Big\rVert ^{2} \\
        & \le \Big\lVert   \boldsymbol{y}^{k}  - \boldsymbol{y}^{\star}  
        - \delta  \Big[ \bar{\boldsymbol{F}}^{\rm W}(\boldsymbol{y}^{k} )
        - \bar{\boldsymbol{F}}^{\rm W}(\boldsymbol{y}^{\star}) \\
        &\quad + \boldsymbol{A} \left( \boldsymbol{\mu}^{k}  
        - \boldsymbol{\mu}^{\star} \right)
        + \boldsymbol{B} \left( \boldsymbol{\lambda}^{k} 
        - \boldsymbol{\lambda}^{\star} \right) \Big] \Big\rVert^{2} \\
        %
        %
        % = & \| \boldsymbol{y}^{k}  - \boldsymbol{y}^{\star} \|^{2} 
        % %
        % + \delta^{2} \Big\lVert \bar{\boldsymbol{F}}^{\rm W}(\boldsymbol{y}^{k} )
        % %
        % - \bar{\boldsymbol{F}}^{\rm W}(\boldsymbol{y}^{\star}) \\
        % %
        % %
        % & + \boldsymbol{A} \left( \boldsymbol{\mu}^{k}  
        % %
        % - \boldsymbol{\mu}^{\star} \right)
        % %
        % + \boldsymbol{B} \left( \boldsymbol{\lambda}^{k} 
        % %
        % - \boldsymbol{\lambda}^{\star} \right)  \Big\rVert^{2}  \\
        % %
        % %
        % & - 2 \delta
        % %
        % ( \boldsymbol{y}^{k}  - \boldsymbol{y}^{\star} )^{\top} 
        % %
        % \Big[ \bar{\boldsymbol{F}}^{\rm W}(\boldsymbol{y}^{k} )
        % %
        % - \bar{\boldsymbol{F}}^{\rm W}(\boldsymbol{y}^{\star})  \\
        % %
        % %
        % & + \boldsymbol{A} \left( \boldsymbol{\mu}^{k}  
        % %
        % - \boldsymbol{\mu}^{\star} \right)
        % %
        % + \boldsymbol{B} \left( \boldsymbol{\lambda}^{k} 
        % %
        % - \boldsymbol{\lambda}^{\star} \right) \Big]  \\
        % %
        % %
        % %
        &\le \| \boldsymbol{y}^{k}  - \boldsymbol{y}^{\star} \|^{2} 
        + 2 \delta^{2} \left\lVert \bar{\boldsymbol{F}}^{\rm W}(\boldsymbol{y}^{k} )
        - \bar{\boldsymbol{F}}^{\rm W}(\boldsymbol{y}^{\star})  \right\rVert^{2}  \\
        & \ + 2 \delta^{2} \left\lVert  \boldsymbol{A} \left( \boldsymbol{\mu}^{k}  
        - \boldsymbol{\mu}^{\star} \right) \right\rVert^{2} 
        + 2\delta^{2} \left\lVert \boldsymbol{B} ( \boldsymbol{\lambda}^{k} 
        - \boldsymbol{\lambda}^{\star} )  \right\rVert^{2} \\
        & \ - 2 \delta
        ( \boldsymbol{y}^{k}  - \boldsymbol{y}^{\star} )^{\top} 
        \left( \bar{\boldsymbol{F}}^{\rm W}(\boldsymbol{y}^{k} )
        - \bar{\boldsymbol{F}}^{\rm W}(\boldsymbol{y}^{\star})  \right) \\
        & \ - 2 \delta
        ( \boldsymbol{y}^{k} \! - \boldsymbol{y}^{\star} )^{\top} 
         \boldsymbol{A} \left( \boldsymbol{\mu}^{k}  
        \! - \boldsymbol{\mu}^{\star} \right)  
        - 2 \delta
        ( \boldsymbol{y}^{k} \! - \boldsymbol{y}^{\star} )^{\top} 
         \boldsymbol{B} ( \boldsymbol{\lambda}^{k} 
        \! - \boldsymbol{\lambda}^{\star}) ,
\end{align*}
where the first inequality is by the nonexpansiveness property of projection operator, and the last inequality is using the identity $(a+b+c)^{2} \le 2a^{2} + 2b^{2} + 2c^{2} $.

By a similar and simple argument, the auxiliary distance
\begin{equation}     \nonumber
    \begin{aligned}
         \| \boldsymbol{\nu}^{k + 1} - \boldsymbol{\nu}^{\star} \|^{2}
        = & \|   \boldsymbol{\nu}^{k} + \delta L \boldsymbol{\mu}^{k} 
        - \boldsymbol{\nu}^{\star} - \delta L \boldsymbol{\mu}^{\star}  \|^{2} \\
        \le &  \|   \boldsymbol{\nu}^{k}  - \boldsymbol{\nu}^{\star} \|^{2}
        + \delta^{2} \|  L (\boldsymbol{\mu}^{k}  -  \boldsymbol{\mu}^{\star})  \|^{2} \\
        & + 2 \delta (\boldsymbol{\nu}^{k}  - \boldsymbol{\nu}^{\star})^{\top}
        L (\boldsymbol{\mu}^{k}  -  \boldsymbol{\mu}^{\star}).
    \end{aligned}
\end{equation}

Using the nonexpansiveness property of the projection operator, we obtain
    \begin{align*}
        & \| \boldsymbol{\mu}^{k + 1} - \boldsymbol{\mu}^{\star} \|^{2} 
        %
        %
        % = & \Big\lVert \Pi_{\mathbb{R}_{+}^{N}}  
        % %
        % \left[ \boldsymbol{\mu}^{k} 
        % %
        % + \delta (\boldsymbol{A}^{\top} \boldsymbol{y}^{k} - \boldsymbol{b}  ) 
        % %
        % - \delta L \boldsymbol{\mu}^{k} - \delta L \boldsymbol{\nu}^{k} \right]  \\
        % %
        % %
        % & -   \Pi_{\mathbb{R}_{+}^{N}}  
        % %
        % \left[ \boldsymbol{\mu}^{\star} 
        % %
        % + \delta (\boldsymbol{A}^{\top} \boldsymbol{y}^{\star} - \boldsymbol{b}  ) 
        % %
        % - \delta L \boldsymbol{\mu}^{\star} - \delta L \boldsymbol{\nu}^{\star} \right] \Big\rVert^{2} \\
        % %
        % %
        \le  \| \boldsymbol{\mu}^{k} - \boldsymbol{\mu}^{\star} \|^{2} 
        + \delta^{2} \| \boldsymbol{A}^{\top}  (\boldsymbol{y}^{k} - \boldsymbol{y}^{\star})\|^{2} \\
        & + \delta^{2} \| L ( \boldsymbol{\mu}^{k} - \boldsymbol{\mu}^{\star} ) \|^{2}  
        + \delta^{2} \| L ( \boldsymbol{\nu}^{k} - \boldsymbol{\nu}^{\star} ) \|^{2}   \\
        & + 2 \delta (\boldsymbol{\mu}^{k} - \boldsymbol{\mu}^{\star})^{\top}
        \boldsymbol{A}^{\top}  (\boldsymbol{y}^{k} \! - \boldsymbol{y}^{\star}) \\
        &- 2 \delta (\boldsymbol{\mu}^{k} \! - \boldsymbol{\mu}^{\star})^{\top}
        L ( \boldsymbol{\mu}^{k}\! - \boldsymbol{\mu}^{\star} ) 
         - 2 \delta (\boldsymbol{\mu}^{k} - \boldsymbol{\mu}^{\star})^{\top}
        L ( \boldsymbol{\nu}^{k} - \boldsymbol{\nu}^{\star} ).
    \end{align*}

It is straightforward to prove that the last term
\begin{equation}   \nonumber
    \begin{aligned}
        \| \boldsymbol{\lambda}^{k + 1} - \boldsymbol{\lambda}^{\star} \|^{2} 
        = & \| \boldsymbol{\lambda}^{k} 
        + \delta \boldsymbol{B}^{\top} \boldsymbol{y}^{k}
        - \boldsymbol{\lambda}^{\star} 
        - \delta \boldsymbol{B}^{\top} \boldsymbol{y}^{\star}
        \| ^{2}    \\
        \le & \| \boldsymbol{\lambda}^{k} - \boldsymbol{\lambda}^{\star} \|^{2}
        + \delta^{2} \|
        \boldsymbol{B}^{\top} ( \boldsymbol{y}^{k} - \boldsymbol{y}^{\star} )
        \| ^{2}   \\
        & + 2 \delta (\boldsymbol{\lambda}^{k} - \boldsymbol{\lambda}^{\star})^{\top}
        \boldsymbol{B}^{\top} ( \boldsymbol{y}^{k} - \boldsymbol{y}^{\star} ) .
    \end{aligned}
\end{equation}
Define $\Delta V_{s}(\chi^{k}) : = V_{s}(\chi^{k + 1}) - V_{s}(\chi^{k})$, combining all these parts, we have
\begin{equation}     \label{eq:lyap_diffe_of_slow_dyna}
    \begin{aligned}
        \Delta V_{s}(\chi^{k}) 
        \le & - 2 \delta
        ( \boldsymbol{y}^{k}  - \boldsymbol{y}^{\star} )^{\top} 
        \left( \bar{\boldsymbol{F}}^{\rm W}(\boldsymbol{y}^{k} )
        - \bar{\boldsymbol{F}}^{\rm W}(\boldsymbol{y}^{\star})  \right)   \\
        & - 2 \delta (\boldsymbol{\mu}^{k} - \boldsymbol{\mu}^{\star})^{\top}
        L ( \boldsymbol{\mu}^{k} - \boldsymbol{\mu}^{\star} ) 
        + \Theta(\delta ^{2}),
    \end{aligned}
\end{equation}
where $\Theta(\cdot)$ denotes a second order function of $\delta$, which is dominated by the first two terms in \eqref{eq:lyap_diffe_of_slow_dyna} for sufficiently small~$\delta$. 

\textbf{Step 6.} Unlike the case in which the Lyapunov function difference for the fast system is required to be negative definite in discrete-time \cite[Th. 2.5]{carnevale2024tracking} and continuous-time \cite[Th. 11.3]{khalil_nonlinear_2002} singularly perturbed systems, \eqref{eq:lyap_diffe_of_slow_dyna} is only negative semidefinite. Now, consider a composite Lyapunov function candidate
\begin{equation}   \nonumber
    V (\chi^{k}, \psi^{k}) : = V_{s} (\chi^{k}) + V_{f} (\psi^{k}).
\end{equation}
By evaluating 
$\Delta V (\chi^{k}, \psi^{k}) : = V (\chi^{k + 1}, \psi^{k + 1})
- V (\chi^{k}, \psi^{k}) 
= \Delta V_{f} (\psi^{k}) + \Delta V_{s}(\chi^{k})$
with combining \eqref{eq:lyap_for_fast_dyna} and \eqref{eq:lyap_diffe_of_slow_dyna}, we have
\begin{equation}   \nonumber
    \begin{aligned}
        \Delta V (\chi^{k}, \psi^{k}) \le 
        & - \| \psi^{k}  \|^{2}  \\
        & - 2 \delta
        ( \boldsymbol{y}^{k}  - \boldsymbol{y}^{\star} )^{\top} 
        \left( \bar{\boldsymbol{F}}^{\rm W}(\boldsymbol{y}^{k} )
        - \bar{\boldsymbol{F}}^{\rm W}(\boldsymbol{y}^{\star})  \right)   \\
        & - 2 \delta (\boldsymbol{\mu}^{k} - \boldsymbol{\mu}^{\star})^{\top}
        L ( \boldsymbol{\mu}^{k} - \boldsymbol{\mu}^{\star} )
        + \Theta(\delta ^{2}).
    \end{aligned}
\end{equation}

Under Assumption \ref{asp:convexity_and_nonempty}, for all $\boldsymbol{y}^{k},\boldsymbol{y}^{\star} \in \boldsymbol{\Omega}$, we first expand the inner product as $( \bar{\boldsymbol{F}}^{\rm W}(\boldsymbol{y}^{k}) - \allowbreak \bar{\boldsymbol{F}}^{\rm W}(\boldsymbol{y}^{\star}) )^{\top} (\boldsymbol{y}^{k} - \boldsymbol{y}^{\star}) = \allowbreak \sum_{i=1}^{N} ( \nabla p_{i}(x_{i}^{k}) - \allowbreak \nabla p_{i}(x_{i}^{\star}) )^{\top} (x_{i}^{k} - x_{i}^{\star}) \allowbreak + N [ q(\sigma(\boldsymbol{x}^{k})) - \allowbreak q(\sigma(\boldsymbol{x}^{\star})) ]^{\top} ( \sigma(\boldsymbol{x}^{k}) - \allowbreak \sigma(\boldsymbol{x}^{\star}) )$. Utilizing the strong convexity parameters $\alpha$ and $\beta$, this expression can be lower bounded by $\sum_{i=1}^{N} \alpha \| x_{i}^{k} - x_{i}^{\star} \|^{2} \allowbreak + N \beta \| \sigma(\boldsymbol{x}^{k}) - \sigma(\boldsymbol{x}^{\star}) \|^{2}$, which is further bounded below by $\alpha \| \boldsymbol{x}^{k} -\boldsymbol{x}^{\star} \|^{2}$.
    
Given the positive semidefinite matrix $L$, we can conclude that the whole singularly perturbed system \eqref{eq:singu_per_sys} is Lyapunov stable. Now, let us investigate the set
\begin{equation}  \label{eq:invariant_set}
    E = \{ (\chi,\psi ): \Delta V(\chi, \psi) = 0  \},
\end{equation}
and let $M$ denote its largest invariant subset. By applying the LaSalle invariance principle \cite[Th. 3.7]{ge2011invariance}, $(\chi^{k}, \psi^{k})$ converges to $M$ as $k \to \infty$. According to \eqref{eq:invariant_set}, we have
\begin{equation}   \nonumber
    E = \{(\boldsymbol{y},\boldsymbol{\nu},\boldsymbol{\mu},\boldsymbol{\lambda}, \psi) :  \boldsymbol{y} = \boldsymbol{y}^{\star},\ L (\boldsymbol{\mu} -\boldsymbol{\mu}^{\star}) = \mathbf{0}_{N}, \ \psi = \psi^{\star}  \}.
\end{equation}

Consider the trajectory $(\bar{\boldsymbol{y}},\bar{\boldsymbol{\nu}},\bar{\boldsymbol{\mu}},\bar{\boldsymbol{\lambda}}, \bar{\psi})$ in $M$, then $\bar{\boldsymbol{y}} \equiv \boldsymbol{y}^{\star}$, $\bar{\psi} \equiv \psi^{\star} = 0$, and $\bar{\boldsymbol{\mu}}$ will remain constant. 
Observing the slow dynamics \eqref{eq:slow_dyna}, we can conclude that $\bar{\boldsymbol{\nu}}$ and $\bar{\boldsymbol{\lambda}}$ will also remain constant. 
Therefore, $(\bar{\boldsymbol{y}},\bar{\boldsymbol{\nu}},\bar{\boldsymbol{\mu}},\bar{\boldsymbol{\lambda}})$ is the equilibrium of \eqref{eq:slow_dyna}, i.e., RGWE $(\boldsymbol{y}^{\star}, \boldsymbol{\nu}^{\star}, \boldsymbol{\mu}^{\star}, \boldsymbol{\lambda}^{\star})$. By LaSalle's invariance principle, \eqref{eq:trans_sin_per_sys} converges to the RGWE $(\boldsymbol{y}^{\star}, \boldsymbol{\nu}^{\star}, \boldsymbol{\mu}^{\star}, \boldsymbol{\lambda}^{\star})$ asymptotically.

Thus, algorithm \eqref{eq:algorithm_compact} converges to an RGWE. This completes the proof.

\subsection{Proof of Theorem \ref{thm:distance}}   \label{appen:C}
According to Proposition \ref{prop:exi_of_WE}, any solution $\boldsymbol{y}^{\rm W}$ of VI($\boldsymbol{\Omega},\bar{\boldsymbol{F}}^{\rm W} $) is an RGWE of the game $\mathcal{G}$. 
    Analogously, the RGNE $\boldsymbol{y}^{\rm N}$ is obtained based on Proposition \ref{prop:exi_of_NE}.
    Therefore, $\boldsymbol{y}^{\rm W}$ and $\boldsymbol{y}^{\rm N}$ satisfy the following variational inequalities:
    \begin{subequations}     \label{eq:VI_for_N_and_W}
        \begin{align}
            \bar{\boldsymbol{F}}^{\rm W}({\boldsymbol{y}^{\rm W})^{\top}
            (\boldsymbol{y}^{1}-\boldsymbol{y}^{\rm W}}) & \ge 0,\ 
            \forall \boldsymbol{y}^{1} \in \boldsymbol{\Omega},  \label{eq:VI_for_W} \\
            \bar{\boldsymbol{F}}^{\rm N}(\boldsymbol{y}^{\rm N})^{\top}
            (\boldsymbol{y}^{2}-\boldsymbol{y}^{\rm N}) & \ge 0,\ 
            \forall \boldsymbol{y}^{2} \in \boldsymbol{\Omega}. \label{eq:VI_for_N}
        \end{align}
    \end{subequations}

     Observe that it is valid to set
     $\boldsymbol{y}^{1}=\boldsymbol{y}^{\rm N}$ in (\ref{eq:VI_for_W}) 
     and $\boldsymbol{y}^{2}=\boldsymbol{y}^{\rm W}$ in (\ref{eq:VI_for_N}). 
     Substituting (\ref{eq:i_gradient_Wardrop}) into (\ref{eq:VI_for_W}) yields
    \begin{align}     \label{eq:Wardrop_dis}
            & \sum^{N}_{i=1}\left( \nabla p_{i}(x^{\rm W}_{i}) 
            + q( \sigma(\boldsymbol{x}^{\rm W}) ) \right)^{\top} 
            ({x}^{\rm N}_{i} - {x}^{\rm W}_{i}) \nonumber\\
            & =\! N \! q( \sigma(\boldsymbol{x}^{\rm W}\!) )^{\! \top} 
            \! (\sigma(\boldsymbol{x}^{\rm N\!}) \! - \! \sigma(\boldsymbol{x}^{\rm W\!})) 
            \! + \! \sum^{N}_{i=1} \!  \left( \nabla p_{i}(x^{\rm W}_{i}) \right)^{\! \top} \! (x^{\rm N}_{i} \! - \! x^{\rm W}_{i}\!) \nonumber  \\
            & \geq 0.
    \end{align}
    Moreover, substituting (\ref{eq:i_gradient_Nash}) into (\ref{eq:VI_for_N}) yields
        \begin{align}
            & \sum^{N}_{i=1} \left( \nabla p_{i} (x^{\rm N}_{i}) 
            + \frac{1}{N}\nabla_{z}q(z)_{|z=\sigma(\boldsymbol{x}^{\rm N})} 
            x^{\rm N}_{i} \right) ^{\top} ({x}^{\rm W}_{i} - x^{\rm N}_{i}) \nonumber \\
            & + \sum^{N}_{i=1}q( \sigma(\boldsymbol{x}^{\rm N}) )^{\top}
            ({x}^{\rm W}_{i} - x^{\rm N}_{i}) \nonumber  \\
            = & Nq( \sigma(\boldsymbol{x}^{\rm N}) ) ^{\! \top} \!
            (\sigma(\boldsymbol{x}^{\rm W})\! - \! \sigma(\boldsymbol{x}^{\rm N})) \nonumber 
            \! + \! \sum^{N}_{i=1} \!  \left(\nabla p_{i}
            (x^{\rm N}_{i})\right)^{\! \top} \! (x^{\rm W}_{i} \! - \! x^{\rm N}_{i}) \nonumber \\
            &  + \frac{1}{N}\sum^{N}_{i=1} 
            \left( \nabla_{z} q(z)_{|z=\sigma(\boldsymbol{x}^{\rm N})} x^{\rm N}_{i} \right)^\top 
            (x^{\rm W}_{i} - x^{\rm N}_{i})
            \geq 0.  \label{eq:Nash_dis}
        \end{align}

    By the $\beta$-strong monotonicity of $q(\cdot)$ on $\mathbb{R}^{n}$ in Assumption~\ref{asp:convexity_and_nonempty},
    \begin{align}    \label{eq:sum}
            & \beta \| \sigma(\boldsymbol{x}^{\rm W}) - \sigma(\boldsymbol{x}^{\rm N}) \| ^{2}  \nonumber \\
            \leq & \left[  q( \sigma(\boldsymbol{x}^{\rm W})) 
            -  q( \sigma(\boldsymbol{x}^{\rm N})) \right] ^{\top} 
            \left( \sigma(\boldsymbol{x}^{\rm W}) - \sigma(\boldsymbol{x}^{\rm N}) \right)  \nonumber \\
            = & q( \sigma(\boldsymbol{x}^{\rm W\!})) ^{\! \top} \!
            \left( \sigma(\boldsymbol{x}^{\rm W\!}) \! - \! \sigma(\boldsymbol{x}^{\rm N}) \right)
            -q( \sigma(\boldsymbol{x}^{\rm N\!})) ^{\! \top}  \!
            \left(\sigma(\boldsymbol{x}^{\rm W\!}) \! - \! \sigma(\boldsymbol{x}^{\rm N\!})\right)  \nonumber \\
           \leq &\! \frac{1}{N} \! \sum^{N}_{i=1}  \!
            \left(\nabla p_{i}(x^{\rm W\!}_{i})\right)^{\! \top} \!
            (x^{\rm N\!}_{i} \! - \! x^{\rm W\!}_{i}) 
            \! - \! q( \sigma(\boldsymbol{x}^{\rm N\!})) 
            ^{\! \top}  \!
            \left(\sigma(\boldsymbol{x}^{\rm W\!}) - \sigma(\boldsymbol{x}^{\rm N\!})\right)  \nonumber \\
            \leq & \! \frac{1}{N}\!  \sum^{N}_{i=1} \! \left( \nabla p_{i}(x^{\rm W\!}_{i}) \right)^{\! \top} \!
            (x^{\rm N\!}_{i} \! - \! x^{\rm W\!}_{i})  
            \! + \! \frac{1}{N} \! \sum^{N}_{i=1} \!
            \left(\nabla p_{i}(x^{\rm N}_{i}) \right)^{\! \top} \!
            (x^{\rm W\!}_{i} \! - \! x^{\rm N\!}_{i})  \nonumber  \\
            & + \frac{1}{N^{2}}\sum^{N}_{i=1} 
            \left( \nabla_{z}q(z)_{|z=\sigma(\boldsymbol{x}^{\rm N})} x^{\rm N}_{i} \right)^\top 
            (x^{\rm W}_{i} - x^{\rm N}_{i}),
    \end{align}
    where we invoke (\ref{eq:Wardrop_dis}) for the first inequality, and (\ref{eq:Nash_dis}) for the second. 
    Since $p_{i}(\cdot)$ is $\alpha$-strongly convex (cf. Assumption \ref{asp:convexity_and_nonempty}), the first two terms of (\ref{eq:sum}), which is related to $p_{i}$, satisfy
    \begin{equation}      \label{eq:strongly_mono_dis}
        \begin{aligned}
            &\! \! \frac{1}{N} \!\sum^{N}_{i=1}  \!
            \left( \nabla p_{i}(x^{\rm N\!}_{i}) \! - \! \nabla p_{i}(x^{\rm W\!}_{i}) \right)^{\! \top} \! \!
            (x^{\rm W\!}_{i} - x^{\rm N\!}_{i}) \!
            \leq \! -  \frac{\alpha }{N} \! \sum^{N}_{i=1} \! \| x^{\rm W\!}_{i} \! - \! x^{\rm N\!}_{i} \| ^{2}\!
        \end{aligned}
    \end{equation}

    As established in \cite[Th. 1.3)]{paccagnan2018nash}, the remaining part of \eqref{eq:sum} satisfies the following inequality:
    \begin{align}       \label{eq:result_from_N_and_W_paper}
             \frac{1}{N^{2}}\sum^{N}_{i=1}  
             \left( \nabla_{z}q(z)_{|z=\sigma(\boldsymbol{x}^{\rm N})} x^{\rm N}_{i} \right)^\top 
             (x^{\rm W}_{i} - x^{\rm N}_{i})  \leq \frac{2R_{x}^{2} L_{q}}{N}.
    \end{align}
    Note that 
    \begin{equation}        \label{eq:Jensen_ineq}
        \begin{aligned}
            & \| \sigma(\boldsymbol{x}^{\rm W}) - \sigma(\boldsymbol{x}^{\rm N}) \| ^{2} 
            = \| \frac{1}{N} \sum^{N}_{i=1} x^{\rm W}_{i} - \frac{1}{N} \sum^{N}_{i=1} x^{\rm N}_{i} \| ^{2} \\
            = & \frac{1}{N^{2}} \| \sum^{N}_{i=1} (x^{\rm W}_{i} - x^{\rm N}_{i}) \| ^{2} 
            \le \frac{1}{N}  \sum^{N}_{i=1} \| x^{\rm W}_{i} - x^{\rm N}_{i} \| ^{2},
        \end{aligned}
    \end{equation}
    where the last inequality is by Jensen's inequality, since $\| \cdot \|^{2}$ is convex on $\mathbb{R}^{n}$, implying 
    $\| \sum_{i=1}^{N} \frac{1}{N} (x^{\rm W}_{i} - x^{\rm N}_{i}) \| ^{2} 
    \le \sum_{i=1}^{N} \frac{1}{N} \| x^{\rm W}_{i} - x^{\rm N}_{i} \| ^{2}$. 
    Substituting (\ref{eq:strongly_mono_dis}) -- (\ref{eq:Jensen_ineq}) into (\ref{eq:sum}) yields
    \begin{align*}
        \beta \| \sigma(\boldsymbol{x}^{\rm W}) - \sigma(\boldsymbol{x}^{\rm N}) \| ^{2}  
        \le & - \frac{\alpha }{N} \sum^{N}_{i=1}  \| x^{\rm W}_{i} - x^{\rm N}_{i} \| ^{2} + \frac{2R_{x}^{2} L_{q}}{N} \\
        \leq & -\alpha  \| \sigma(\boldsymbol{x}^{\rm W}) - \sigma(\boldsymbol{x}^{\rm N}) \| ^{2} 
        + \frac{2R_{x}^{2} L_{q}}{N}.
    \end{align*}
    This simplifies to
    \begin{equation}      \label{eq:distance_of_sigma}
        \| \sigma(\boldsymbol{x}^{\rm W}) - \sigma(\boldsymbol{x}^{\rm N}) \| ^{2} 
        \leq \frac{2R_{x}^{2} L_{q}}{N(\beta + \alpha )}.
    \end{equation}
    
    By this critical result (\ref{eq:distance_of_sigma}), we can induce the relationship between $\varepsilon$-RGNE and RGWE. Based on the definition of $\varepsilon$-RGNE (similar to Definition \ref{def:GNE}), for each player $i \in \mathcal{I}$ and for all $x_{i}\in \mathcal{C}_{i,\boldsymbol{\rho}}(\boldsymbol{x}_{-i})$, we have
    \begin{align}   \label{eq:three_parts}
        & \quad \ J_{i}(x^{\rm W}_{i},\sigma(\boldsymbol{x}^{\rm W}))  
        - J_{i}\left(x_{i},\frac{1}{N}  \left(  
        \sum_{j=1,j \neq i}^{N} x_{j}^{\rm W} +  x_{i}  \right) \right) \nonumber \\
        & = \underbrace{J_{i}(x^{\rm W}_{i},\sigma(\boldsymbol{x}^{\rm W}))
        - J_{i}(x^{\rm N}_{i},\sigma(\boldsymbol{x}^{\rm N}))}_{\text{Term 1}} \nonumber \\
        &\quad + \underbrace{J_{i}(x^{\rm N}_{i},\sigma(\boldsymbol{x}^{\rm N}))  
        - J_{i}\left(x_{i}, \frac{1}{N} \left( \sum_{j=1,j\neq i}^{N}x_{j}^{\rm N} + x_{i} \right) \right)}_{\text{Term 2}} \nonumber \\
        &\quad + \underbrace{%
        J_{i}\left(x_{i}, \frac{1}{N} \left( \sum_{\substack{j=1, j\neq i}}^{N} x_{j}^{\rm N} + x_{i} \right) \right)
        }_{\text{Term 3 (part 1)}} \nonumber \\
        &\quad -\underbrace{%
        J_{i}\left(x_{i}, \frac{1}{N} \left( \sum_{\substack{j=1, j\neq i}}^{N} x_{j}^{\rm W} + x_{i} \right) \right)
        }_{\text{Term 3 (part 2)}}.
    \end{align}
    Replaying the technique above yields
    \begin{equation}      \nonumber
    \begin{aligned}
        & J_{i}(x^{\rm W}_{i},\sigma(\boldsymbol{x}^{\rm W})) 
        - J_{i}(x^{\rm N}_{i},\sigma(\boldsymbol{x}^{\rm N})) \\
        = & J_{i}(x^{\rm W}_{i},\sigma(\boldsymbol{x}^{\rm W})) 
        - J_{i}(x^{\rm N}_{i},\sigma(\boldsymbol{x}^{\rm W})) 
        + J_{i}(x^{\rm N}_{i},\sigma(\boldsymbol{x}^{\rm W}))   \\
        &  - J_{i}(x^{\rm N}_{i},\sigma(\boldsymbol{x}^{\rm N})) \\
        \le & J_{i}(x^{\rm N}_{i},\sigma(\boldsymbol{x}^{\rm W}))  
        - J_{i}(x^{\rm N}_{i},\sigma(\boldsymbol{x}^{\rm N})),
    \end{aligned}
    \end{equation}
    where the inequality comes from the definition of RGWE. By Assumption \ref{asp:convexity_and_nonempty}, the inequality above could be written as 
    \begin{align*}   
        & J_{i}(x^{\rm W}_{i},\sigma(\boldsymbol{x}^{\rm W}))  
        - J_{i}(x^{\rm N}_{i},\sigma(\boldsymbol{x}^{\rm N}))  \\
        \le & J_{i}(x^{\rm N}_{i},\sigma(\boldsymbol{x}^{\rm W})) 
        - J_{i}(x^{\rm N}_{i},\sigma(\boldsymbol{x}^{\rm N})) \\
        \le & \| p_{i}(x^{\rm N}_{i}) + q(\sigma(\boldsymbol{x}^{\rm W}))^{\top}x^{\rm N}_{i} 
        - p_{i}(x^{\rm N}_{i}) - q(\sigma(\boldsymbol{x}^{\rm N}))^{\top}x^{\rm N}_{i}  \|  \\
        \le & \|  q(\sigma(\boldsymbol{x}^{\rm W})) - 
        q(\sigma(\boldsymbol{x}^{\rm N})) \| \ \|  x^{\rm N}_{i}  \|  \\
        \le & R_{x} L_{q}  \| \sigma(\boldsymbol{x}^{\rm W}) - \sigma(\boldsymbol{x}^{\rm N})   \|,
    \end{align*}
    where the third inequality follows from the Cauchy-Schwarz inequality, and the last inequality is by the $L_{q}$-Lipschitz continuity of $q(\cdot)$.
   Term 2 is exactly the definition of RGNE, implying 
    $J_{i}(x^{\rm N}_{i},\sigma(\boldsymbol{x}^{\rm N}))  
    - J_{i}\left(x_{i}, \frac{1}{N} \left( \sum_{j=1,j\neq i}^{N}x_{j}^{\rm N} + x_{i} \right) \right)\le 0$, for all $x_{i}\in \mathcal{C}_{i,\boldsymbol{\rho}}(\boldsymbol{x}_{-i})$. 
    Next, we consider Term 3, which has a similar form to Term 1. 
    Notice that 
    $$\max_{i \in \mathcal{I}} \max_{x_{i} \in \mathcal{C}_{i,\boldsymbol{\rho}}(\boldsymbol{x}_{-i}) \subseteq \mathcal{X}_{i} } \| x_{i} \| \le \max_{i \in \mathcal{I}} \max_{x_{i} \in \mathcal{X}_{i} } \| x_{i} \| = R_{x}.$$
    Thus, based on the above formulation, we have
    \begin{equation}   \nonumber
    \begin{aligned}
        \text{Term 3}  = & J_{i}(x_{i},\sigma(\boldsymbol{x}^{\rm N}) + \frac{1}{N} x_{i} - \frac{1}{N} x_{i}^{\rm N})  \\
        & - J_{i}(x_{i},\sigma(\boldsymbol{x}^{\rm W}) + \frac{1}{N} x_{i} - \frac{1}{N} x_{i}^{\rm W}) \\
        \le & R_{x} L_{q}\| \sigma(\boldsymbol{x}^{\rm W}) - 
        \sigma(\boldsymbol{x}^{\rm N}) 
        -\frac{1}{N} (x_{i}^{\rm N} - x_{i}^{\rm W}) \|  \\
        \le & R_{x} L_{q} \| \sigma(\boldsymbol{x}^{\rm W}) - 
        \sigma(\boldsymbol{x}^{\rm N}) \|
        + \frac{2R^{2}_{x} L_{q}}{N}.
    \end{aligned}
    \end{equation}
    Combining the estimates for Terms 1-3, we obtain
    \begin{align*}  
       & J_{i}(x^{\rm W}_{i},\sigma(\boldsymbol{x}^{\rm W})) 
       - J_{i}\left(x_{i},\frac{1}{N}  \left(  
        \sum_{j=1,j \neq i}^{N} x_{j}^{\rm W} +  x_{i}  \right) \right)  \\
       \le & 2R_{x} L_{q}\| \sigma(\boldsymbol{x}^{\rm W}) - 
       \sigma(\boldsymbol{x}^{\rm N})  \| 
       + \frac{2 R_{x}^{2} L_{q} }{N}   \\
       \leq & 2 R_{x}^{2} L_{q}  \left(
       \sqrt{\frac{2  L_{q}}{N(\alpha + \beta)}} + \frac{1}{N} \right),
    \end{align*}
   where the last inequality is by \eqref{eq:distance_of_sigma}. This completes the proof.

\bibliographystyle{IEEEtran}  
\bibliography{references}

\end{document}